\documentclass[11pt,a4paper]{article}

\usepackage{appendix}

\usepackage{graphicx}
\usepackage{epsfig}

\usepackage{hyperref}
\usepackage{graphicx,epsf}
\usepackage{bm}
\usepackage{amsmath}
\usepackage{amssymb}

\usepackage[dvipsnames]{xcolor}

\newcommand{\vecb}[1]{{\bf #1}}
\def \d {\mathrm{d}}

\begin{document}

\begin{center}
{\large{\textbf{Gyrokinetic investigation of Alfv\'en instabilities in the presence of turbulence}}}\\
\vspace{0.2 cm}
{\normalsize {A. Biancalani$^1$, A. Bottino$^1$, A. Di Siena$^2$, \"O. G\"urcan$^3$, T. Hayward-Schneider$^1$, F. Jenko$^1$, P.~Lauber$^1$, A. Mishchenko$^4$, P. Morel$^3$, I. Novikau$^1$, F. Vannini$^1$,  L. Villard$^5$, and A. Zocco$^4$\\}}
\vspace{0.2 cm}
\small{$^1$ Max Planck Institute for Plasma Physics, 85748 Garching, Germany\\
$^2$ The University of Texas at Austin, 201 E 24th St, 78712 Austin, Texas, USA\\
$^3$ Laboratoire de Physique des Plasmas, CNRS, Ecole Polytechnique, Sorbonne Universit\'e, Universit\'e Paris-Saclay, Observatoire de Paris, F-91120 Palaiseau, France\\
$^4$ Max Planck Institute for Plasma Physics, 17491 Greifswald, Germany\\
$^5$ Ecole Polytechnique F\'ed\'erale de Lausanne, Swiss Plasma Center, CH-1015 Lausanne, Switzerland}\\
\vspace{0.1 cm}
{\footnotesize{contact of main author: \url{http://www2.ipp.mpg.de/~biancala/}}}
\end{center}

\begin{abstract}
The nonlinear dynamics of beta-induced Alfv\'en Eigenmodes (BAE) driven by energetic particles (EP) in the presence of ion-temperature-gradient (ITG) turbulence is investigated, by means of selfconsistent global gyrokinetic simulations and analytical theory. A tokamak magnetic equilibrium with large aspect ratio and reversed shear is considered. A previous study of this configuration has shown that the electron species plays an important role in determining the nonlinear saturation level of a BAE in the absence of turbulence [A. Biancalani, et al., {\it J. Plasma Phys.} (2020)]. Here, we extend the study to a turbulent plasma. The EPs are found modify the heat fluxes by introducing energy at the large spatial scales, mainly at the toroidal mode number of the dominant BAE and its harmonics. In this regime, BAEs are found to carry a strong electron heat flux. The feed-back of the global relaxation of the temperature profiles induced by the BAE, and on the turbulence dynamics, is also discussed.
\end{abstract}


\section{Introduction}
\label{sec:intro}

Magnetically confined plasmas are complex systems in which waves and instabilities at multiple spatial scales coexist and influence each other.
Important examples are microinstabilities, meso-scale zonal flows and macroscopic MHD instabilities like Alfv\'en modes. 
Microinstabilities, like ion-temperature-gradient (ITG) modes~\cite{Rudakov65}, are linearly unstable due to the gradients of plasma temperature, and nonlinearly interact and saturate forming turbulence states. ITGs carry particle and heat fluxes in the direction of the nonuniformity, and therefore they are particularly deleterious to the heat and particle confinement. One of the products of the nonlinear interaction of microinstabilities is the formation of zonal flows (ZF).
ZFs are ExB flows (primarily in the poloidal direction) associated with purely radial variations of the electrostatic potential.
They can play the role of the dominant turbulence saturation mechanism~\cite{Hasegawa79}, by breaking the turbulence vortices, and consequently pushing the energy towards higher radial wavenumbers, where the plasma absorption occurs at kinetic scales.
Alfv\'en modes (AM) are eigenmodes of the shear Alfv\'en waves, i.e. electromagnetic plasma waves propagating as transverse waves along the magnetic field lines.
Various types of AMs exist, such as Global Alfv\'en Eigenmodes (GAE)~\cite{Appert82}, Toroidicity-induced Alfv\'en Eigenmodes (TAE)~\cite{Cheng85} or Beta-induced Alfv\'en Eigenmodes (BAE)~\cite{Chu92,Heibrink93,Zonca96}. AMs can be driven unstable by suprathermal ions, named here energetic particles (EP), present in tokamak plasmas due to external heating mechanisms and to nuclear fusion reactions.
AMs can then lead to a redistribution of the EP population~\cite{Chen16}, which can have consequences, inter alia, on plasma heating.

The difference in temporal and spatial scales has been invoked in the past to justify a separate treatment of AMs and microturbulence. Nevertheless, they can mutually interact either due to direct coupling via wave-wave nonlinear interaction, or by indirect interaction. The wave-wave nonlinear interaction has been proposed to explain the experimental measurements in ASDEX Upgrade~\cite{Maraschek97}, and studied theoretically to investigate the AM saturation~\cite{Chen98}. Regarding the indirect interaction, the reader can see Ref.~\cite{Ida20} for a review of some recent experimental evidences. An indirect interaction can occur for example due to the ZFs, which can be nonlinearly excited by both microinstabilities via parametric excitation~\cite{Hasegawa79,Chen00} and AMs via forced-driven excitation~\cite{Chen12,Qiu16PoP,Qiu16NuFu} (for a theoretical introduction to this interaction mechanism, see Ref.~\cite{Zonca15}). 
In addition, toroidal symmetric structures can be produced in the plasma equilibrium profiles, e.g., density and temperature, which are typically linearly stable and characterized by a slow time variation  with respect to micro-instabilities~\cite{Diamond05}.
The possibility of defining plasma nonlinear equilibria, consistent with zonal structures and microinstabilities was suggested in Refs.~\cite{Chen07,Falessi19} by extracting the part of the toroidally symmetric distribution function that is undamped by collisionless processes.
Another way of indirect interaction is by means of the EP population. EPs are known to excite and nonlinearly interact with AMs~\cite{Chen16}. The interaction of turbulence and EPs has also been observed in tokamak plasma experiments~\cite{Tardini07, Heidbrink09, Romanelli10, Bock17} and investigated by means of analytical theory~\cite{White89, Chen16, Zonca15, Qiu16PoP, Qiu16NuFu} and flux-tube numerical simulations (see for example Ref.~\cite{Angioni09, Zhang10, Holland12, Citrin13, Garcia15, DiSiena18, DiSiena19}).
Finally, a third way of indirect interaction is the nonlinear modification of the plasma profiles (see Ref.~\cite{Zonca15}), as both AMs and microinstabilities can carry heat fluxes. In this paper, we investigate in particular the heat flux carried by AMs and discuss this third indirect interaction mechanism.

Traditionally, global fully gyrokinetic (GK) nonlinear simulations of AMs have been unpractical due to the large computational costs. By global, we mean considering the whole radial domain and the associated variation of the equilibrium plasma profiles. By fully GK, we mean here treating all the plasma species with a GK model.
Consequently, the nonlinear dynamics of AMs has been studied in the past mostly in the absence of turbulence, and with hybrid models treating the EPs and the thermal plasma (or part of it) with GK and fluid models, respectively.
Alternatively, by using local (i.e., flux-tube) models, fully GK simulations of AMs in the presence of turbulence have also been performed, focusing on AMs in the limit of high toroidal mode number~\cite{Bass10,DiSiena19}.
Recently, global, fully GK simulations have become affordable due to the availability of more powerful supercomputers, and more efficient numerical schemes (see for example Ref.~\cite{Cole17}).
In this paper, we present the results of global, fully GK simulations describing the self-consistent nonlinear interaction of AMs and ITG turbulence.

We consider here an analytical magnetic equilibrium with concentric circular flux surfaces, reversed magnetic shear, and large aspect ratio. 
The stability of AMs in this equilibrium has recently been investigated by focusing on the toroidal mode numbers $0\le n \le 9$~\cite{Biancalani20JPP}.
In this paper, we extend the previous study allowing higher-$n$ ITG modes to develop in the same equilibrium, and we study the self-consistent nonlinear interaction.
The main numerical tool used here is the GK particle-in-cell code ORB5. ORB5 was originally written for electrostatic turbulence studies~\cite{Jolliet07}, and then extended to its electromagnetic, multi-species, version~\cite{Bottino11,Lanti20}. ORB5 is based on a variational formulation of the electromagnetic gyrokinetic theory, which ensures appropriate conservation laws~\cite{Tronko19}. It uses state-of-the-art numerical schemes~\cite{Lanti20} that allow for transport time scales simulations.
The global character of ORB5, i.e., the resolution of the full radial extension of the global eigenmodes to scales comparable with the minor radius, makes ORB5 appropriate for studying low-$n$ AMs, without pushing towards the local limit of vanishing ratios of the ion Larmor radius to the tokamak minor radius.
ORB5 has been verified and benchmarked against the linear and nonlinear dynamics of AMs~\cite{Cole17,Koenies18,Taimourzadeh19}, ZFs~\cite{Biancalani14,Biancalani17pop}, and ITG modes~\cite{Goerler16,Tronko17}.
In this paper, we discuss the dynamics of AMs in the presence of ITG turbulence, as shown by ORB5 global self-consistent simulations~\cite{Biancalani19EPS}, and compare with the estimation of analytical theory.

The structure of the paper is the following. The description of the numerical model of ORB5, of the numerical experiment, and of the considered tokamak case, are given in Sec.~\ref{sec:model} and Sec.~\ref{sec:equil-profs}. The heat flux of ITGs and BAEs separately, is described respectively in Sec.~\ref{sec:ITG-lin} and Sec.~\ref{sec:SAW-heat-flux}. The results of the numerical simulations of AMs and turbulence are shown in Sec.~\ref{sec:EM-turb-AMs}, with dedicated subsections on the evolution of the zonal and nonzonal electric fields, on the evolution of the heat fluxes, on the evolution of the temperature profiles, and on the consequent modification of the spectra at low-$n$, dominated by the BAEs and at high-$n$, dominated by the ITGs.
The difference between the heat fluxes carried by ITGs and BAEs is also evaluated analytically, to support the numerical findings, and is described in Sec.~\ref{sec:anal}. Finally, a summary of the results and a discussion are given in Sec.~\ref{sec:conclusions}.


\section{The model}
\label{sec:model}

The investigation of the dynamics of AMs, EPs, and turbulence, requires to treat all 3 species (thermal ions, thermal electrons, and energetic ions) with a kinetic model. Among the main reasons, we mention here the importance of the wave-particle resonances which are essential for determining the ion and electron Landau damping, and for the AM drive due to the EPs.
Due to the low frequencies of the modes of interest, with respect to the ion cyclotron frequency, we are allowed to reduce the complexity of the kinetic model from 6D in phase space to 5D, by averaging out the fast cyclotron motion of the particles around the magnetic field lines: this reduced model is called gyrokinetics (GK).
Global simulations allow to investigate the intrinsically multiscale dynamics of a system made of microturbulence, meso-scale zonal flows, and macro-scale AMs. By means of global simulations, we can investigate the interplay of the several instabilities at the different positions where they develop.

In this paper, we use the GK framework for solving numerical simulations, and for making comparisons with analytical theory, which helps interpreting the results of the simulations.
The main numerical tool used for the investigations described here is the global GK particle-in-cell code ORB5.
The model equations of ORB5 are the gyrocenter trajectories, and the two equations for the fields.

The gyrocenter trajectories are:
\begin{eqnarray}
\dot{\vecb  R}&=&\frac{1}{m}\left(p_\|-\frac{e}{c} \langle A_\parallel\rangle_G \right)\frac{\vecb{B^*}}{B^*_\parallel} + \frac{c}{e B^*_\parallel} \vecb{b}\times \left[\mu
  \nabla B + e \nabla \langle  \phi -  \frac{p_\|}{mc} A_\| \rangle_G \right] \label{eq:trajectories_a} \\
\dot{p_\|}&=&-\frac{\vecb{B^*}}{B^*_\parallel}\cdot\left[\mu \nabla B + e
  \nabla \langle  \phi -  \frac{p_\|}{mc} A_\| \rangle_G \right] \label{eq:trajectories_b}
\end{eqnarray}
The phase-space coordinates are $\vecb{Z}=(\vecb{R},p_\|,\mu)$, i.e. respectively the gyrocenter position, canonical parallel moment $p_\| = m U + (e/c)\langle A_\parallel\rangle_G$ and magnetic moment $\mu = m v_\perp^2 / (2B)$.
The time-dependent fields are named $\phi$ and $A_\|$, and they are respectively the perturbed scalar potential and the parallel component of the perturbed vector potential.
In our notation, on the other hand, $\vecb{A}$ is the equilibrium vector potential. The Jacobian is given by the parallel component of $\vecb{B}^*= \vecb{B} + (c/e) p_\| \vecb{\nabla}\times \vecb{b}$, where $\vecb{B}$ and $\vecb{b}$ are the equilibrium magnetic field and magnetic unit vector. The summation is over all species present in the plasma. The gyroaverage operator is
defined by:
\begin{equation}
 \langle \phi \rangle_G = \frac{1}{2\pi} \int_0^{2\pi} \phi (\vecb{R} + \vecb{\rho}_L) \, \d \alpha
\end{equation}
where $\alpha$ here is the gyroangle and $\rho_L=\rho_L(\alpha,\mu)$ is the Larmor radius. The gyroaverage operator reduces to the zeroth Bessel function $J_0(k_\perp \rho_{Li})$ if we transform into Fourier space. The gyroaverage is calculated for all ion species. For electrons, $\rho_L \rightarrow 0$, therefore $\langle \phi \rangle_G = \phi(\vecb{R})$ (see \cite{Lanti20} for more detail). 
In other words, we take into account finite Larmor radius of ions, and we neglect them for electrons.

The quasineutrality equation is:
\begin{equation}
 -\Sigma_{\rm{sp}\ne e} \vecb{\nabla} \cdot \frac{mc^2\int \d W f_M}{B^2} \nabla_\perp \phi=\Sigma_{\rm{sp}} \int \d W  e \langle f \rangle_G \label{eq:Poisson}
\end{equation}
Here $f$ and $f_M$ are the total and equilibrium (i.e. independent of time) distribution functions, the integrals are over the phase space volume, with $\d V$ being the real-space infinitesimal and $\d W = (2\pi/m^2) B_\|^* \d p_\| \d \mu$ the velocity-space infinitesimal.

Finally, the Amp\`ere equation is:
\begin{eqnarray}
\Sigma_{\rm{sp}} \int \d W \Big( \frac{ep_\|}{mc} f-\frac{e^2}{mc^2} \langle A_\parallel \rangle_G f_{M}
 \Big)  +  \frac{1}{4\pi}\nabla_\perp ^2 A_\parallel =0 \label{eq:Ampere}
\end{eqnarray}

Electromagnetic particle-in-cell gyrokinetic codes are affected by a numerical problem called the ``cancellation problem'', which arises from the fact that some terms in the Amp\`ere equation are solved analytically, and some terms are evaluated by means of a marker discretisation, which introduces a statistical error.
A detailed description of the cancellation problem and possible mitigation techniques can be found in \cite{Mishchenko17}. In particular, the ``pull-back'' mitigation technique used for the simulations discussed in this paper, is described in \cite{Mishchenko18}.

This study of AMs in the presence of turbulence is performed by running a first simulation with turbulence only, and switching on the EP effects in a second simulation, namely a restart. This is done in the following way:\\
- firstly, a simulation with 3 species is initialized, where the EP density is  $\langle n_{EP} \rangle /n_e = 0.01$, and the EP profiles are identical to those of the thermal species;\\
- secondly, a ``restart'' simulation is performed, i.e. starting from the last time step of the previous one, and with modified EP profiles taken from Ref.~\cite{Biancalani20JPP}. In particular, we consider a flat EP temperature profile, with the EP temperature 10 times higher than the thermal species at the reference position at mid-radius. The EP density gradient is 10 times higher than the thermal gradients at the reference position. Moreover, EPs are allowed to redistribute in phase space, i.e. they are allowed to follow perturbed orbits, like the thermal species. This defines fully nonlinear simulations.

The quasineutraility is imposed by ORB5 at every restart, by modifying the electron profile in order to have $n_e=n_i+n_{EP}$ at each radial position. As a consequence, due to the larger density gradient of the EPs species after the ``switch'', the electron density is slightly steeper. The effect of this ``switch'' on the linear dynamics of the ITGs has been found to be negligible.

%

Perturbed modes with toroidal mode number in the range $20 \leq n \leq 30$ are initialized, while all toroidal modes in the range $0 \leq n \leq 40$ are simulated. For each toroidal mode $n$, a radial dependent filter is applied, allowing for poloidal mode numbers in the range $m=n \, q(s) \pm \delta m$.
A spatial grid of (ns, nchi, nphi) = (256, 384, 192) points, a time step of dt=5 $\Omega_i^{-1}$ ($\Omega_i$ being the ion cyclotron frequency), and a number of markers of $(n_i, n_e, n_{EP})$ = (2, 10, 2)e8 respectively for the thermal ions, electrons and EP are used for the turbulence simulations. Unicity boundary conditions are applied at potentials at the axis and Dirichlet at the edge.
No collisions are considered here.
We consider a Krook operator which acts like a source for the thermal species, by restoring the initial plasma profiles, and at the same imposes an artificial damping to all modes, with value $\gamma_K=1.0\cdot 10^{-4} \Omega_i$.

\section{Magnetic equilibrium and plasma profiles}
\label{sec:equil-profs}

\begin{figure}[b!]
\begin{center}
\includegraphics[width=0.45\textwidth]{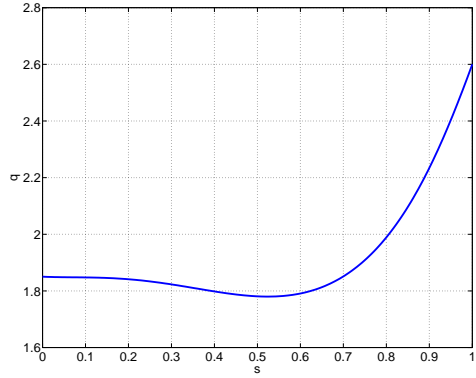}
\vskip -1em
\caption{\label{fig:q_s} Safety factor profile.}
\end{center}
\end{figure}

\begin{figure}[b!]
\begin{center}
\includegraphics[width=0.45\textwidth]{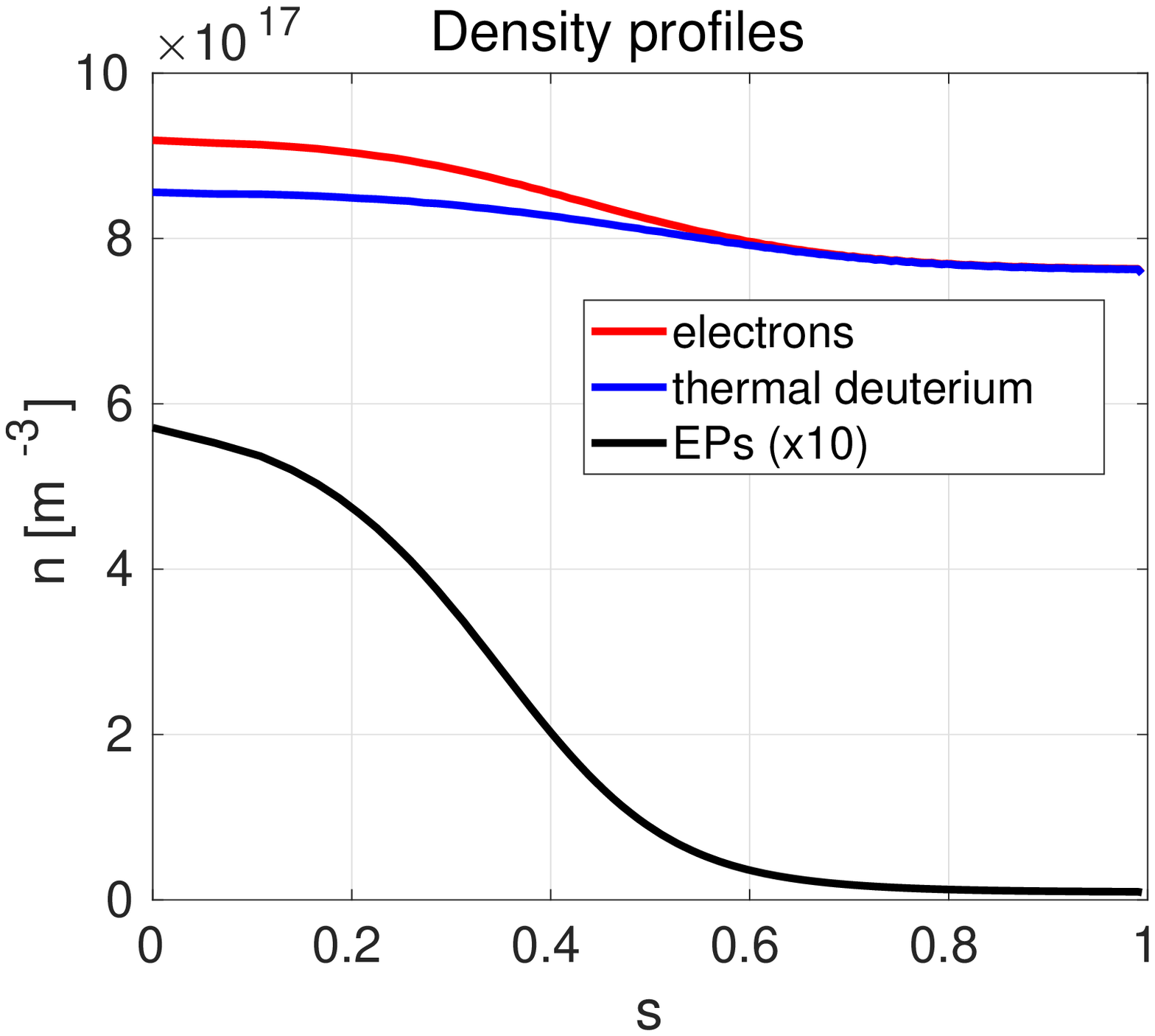}
\includegraphics[width=0.45\textwidth]{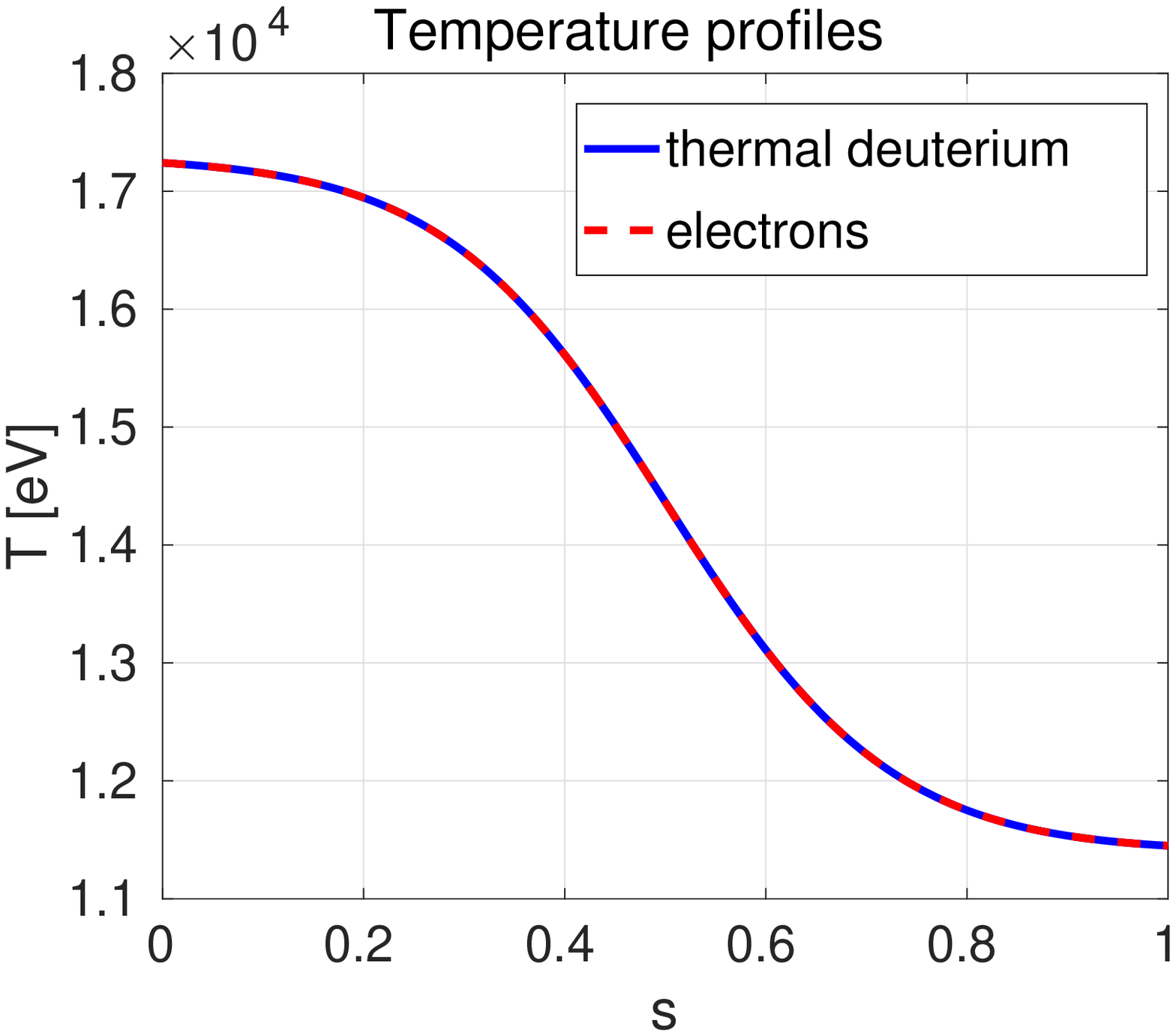}
\vskip -1em
\caption{\label{fig:profs_s} Density (left) and temperature (right) profiles, vs $s$ radial coordinate.}
\end{center} 
\end{figure}

The magnetic equilibrium and plasma profiles used here are the same as in Ref.~\cite{Biancalani20JPP} for the study of AMs,  and in Ref.~\cite{Biancalani19EPS} for the study of AMs and turbulence. The major and minor radii are $R_0 = 1.0$ m, and $a=0.1$ m, and the toroidal magnetic field at the axis is $B_0 = 3.0$ T. An equilibrium with circular concentric flux surfaces is considered. The safety factor has a value of 1.85 at the axis, it decreases from $\rho$=0 to $\rho$=0.5, where the minimum value is located ($q(\rho=0.5)$=1.78), and then it raises to the edge, where it reaches the maximum value ($q(\rho=1)$=2.6) (see Fig.~\ref{fig:q_s}). Here $\rho$ is a normalized radial coordinate defined as $\rho=r/a$.
A reference radial position is chosen at $\rho=\rho_r=0.5$, corresponding to $s=0.525$, where the flux radial coordinate $s$ is defined as $s=\sqrt{\psi_{pol}/\psi_{pol}(edge)}$, $\psi_{pol}$ being the poloidal magnetic flux. The ion and electron temperature profiles are the same: $T_e(\rho)=T_i(\rho)$. The temperature at the reference radius is chosen to have a value of $T_e(\rho=\rho_r)$ corresponding $\rho^* = \rho_s/a = 0.00571$ (with $\rho_s = \sqrt{T_e/m_i}/\Omega_i$ being the sound Larmor radius).
The electron thermal to magnetic pressure ratio of $\beta_e = 8\pi \langle n_e \rangle T_e(\rho_r)/B_0^2 = 5\cdot 10^{-4}$.

The ion species (thermal and energetic) have the mass of the deuterium.
The value of the electron mass is chosen as $m_i/m_e=200$. This value is chosen in order to have quicker numerical simulations. This value is found to be at convergence for the saturation levels of the BAE, and for the linear dynamics of the ITG. The dynamics of the BAE at very low EP concentration depends on the value of the electron mass due to the electron Landau damping, and in particular the BAE in the absence of EPs is found to be slightly above marginal stability, with a growth rate which is small but finite. In this paper, we show the results of simulation where a krook operator is used, which adds an artificial damping to all modes, with a value of $\gamma_K=1.0\cdot 10^{-4} \Omega_i$. This value is sufficiently high to make BAEs linearly stable in the absence of EPs (independently on the electron mass).

The equilibrium considered has some simplifications with respect to present-day tokamaks: namely a higher aspect ratio, and circular concentric flux surfaces. This allows to capture the physics of a simple case, before going to more accurate experimental cases. At the same time, an effort of comparing the results of the code ORB5 with experimental measurements of EP-driven modes is being done in separate papers (\cite{Novikau20,Vannini20}).


\section{ITG linear dynamics}
\label{sec:ITG-lin}

The radial location is near the $s=0.525$ reference surface, where the peak of the temperature gradient is.
The EPs are found not to be sensibly changing the ITG dynamics for concentration lower than $n_{EP}/n_e=0.05$, which is the regime of interest in this paper.

\begin{figure}[t!]
\begin{center}
\includegraphics[width=0.4\textwidth]{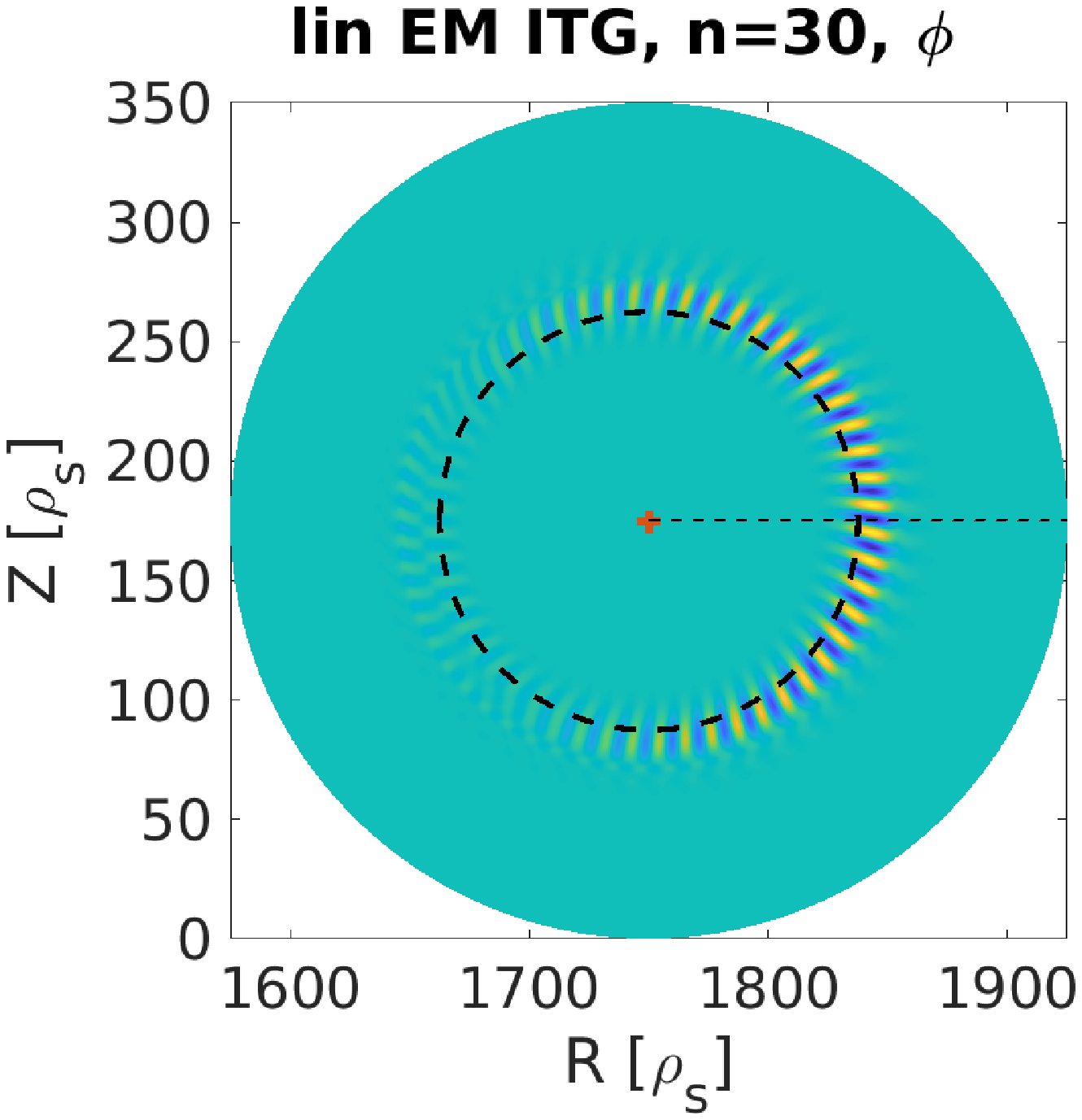}
\includegraphics[width=0.4\textwidth]{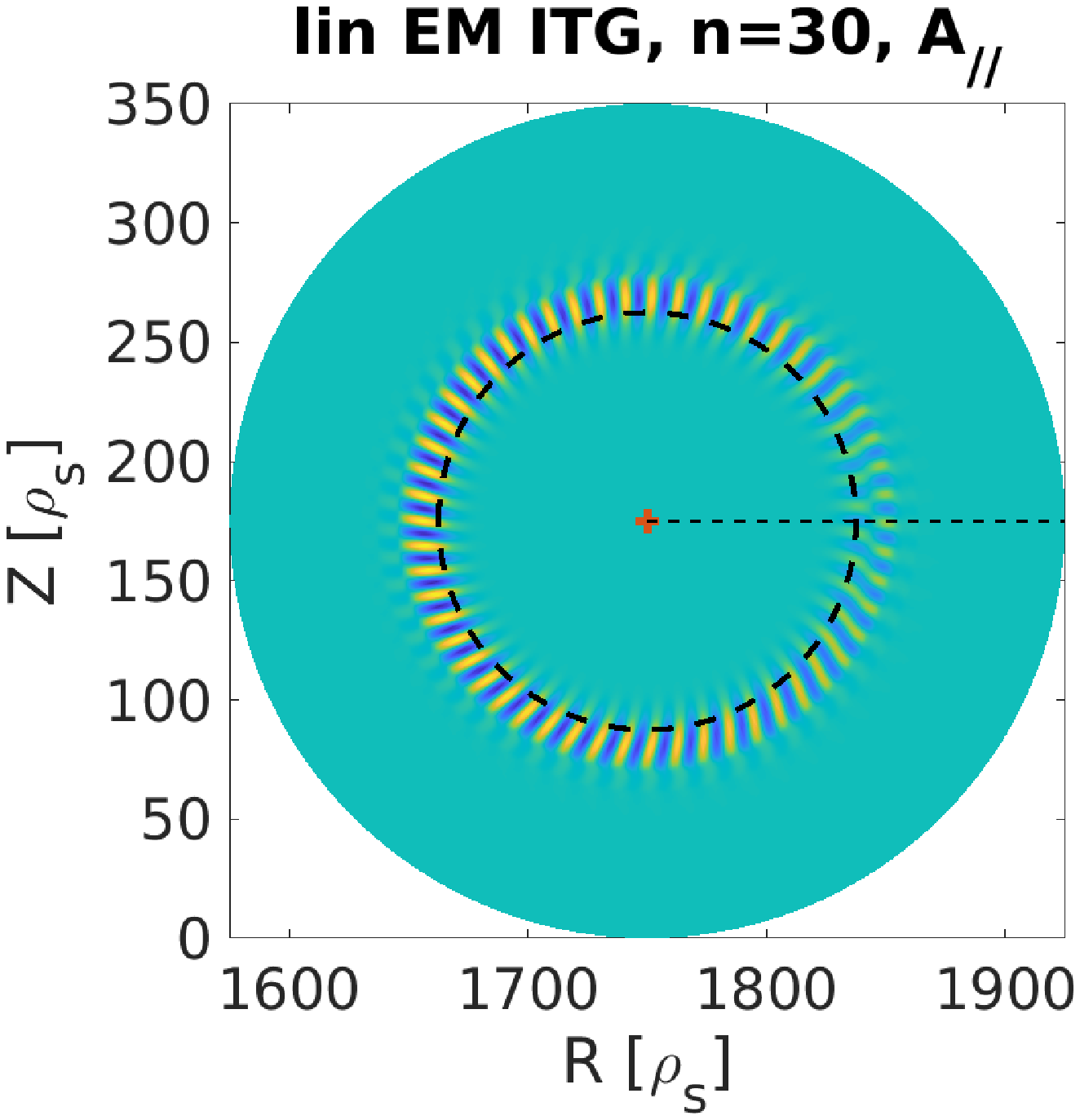}
\caption{\label{fig:ITG-em-structure}  Spatial structure of the scalar potential (left) and parallel component of the vector potential (right) for the linear ITG mode with $n=30$, with radial position $s=0.5$ depicted as a dashed circle. No Krook operator (sources/sinks) is applied here.}
\end{center} 
\end{figure}
Here we show the spectrum of linear electromagnetic simulations of ITG, without sources/sinks, i.e. without krook operator. The most unstable mode is around $n=30$, with growth rate:
\begin{equation}
\gamma_{ITG,EM} = 4.25 \cdot 10^{-4} \; \Omega_i
\end{equation}

%
%
%

\begin{figure}[t!]
\begin{center}
\includegraphics[width=0.45\textwidth]{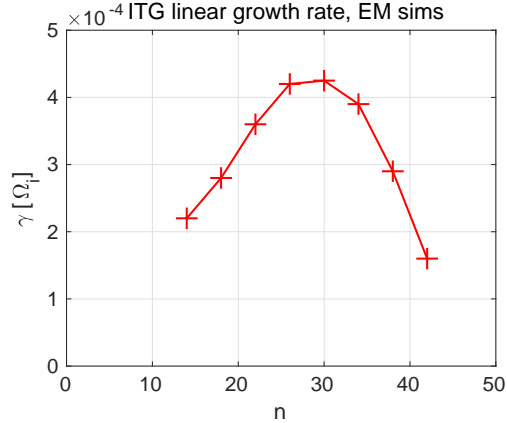}
\caption{\label{fig:ITG-em} Growth rate of linear electromagnetic simulations of ITGs, vs toroidal mode number. No Krook operator (sources/sinks) is applied here.}
\end{center}
\end{figure}

In order to characterise the ITG heat flux, a linear electromagnetic simulation of a single ITG, with $n=26$, is considered. The EP concentration is $n_{EP}/n_e=0.01$
The heat flux is measured, and it is found to grow exponentially. The growth rate of the heat flux is measured to be $\gamma_{h}=0.83e-3$.

We write the volume averaged heat flux as:
\begin{equation}\label{eq:heat-flux-ITG-lin}
\langle \Gamma_s \rangle = \alpha_s \; n^2\; \langle \phi^2 \rangle
\end{equation}
where $s$ is the species index, and $n$ is the toroidal mode number.
We measure for this ITG with $n=26$ the following values respectively for the thermal ions, electrons, and EPs:
\begin{eqnarray}
\alpha_{ITG,i} & = & 0.037\\
\alpha_{ITG,e} & = & 0.014\\
\alpha_{ITG,EP} & = & 0.002
\end{eqnarray}
where $\alpha$ is in units of $n_s T_s c_s \rho^* T_e^2/e^2$, with $n_s$ being the density, and $T_s$ being the temperature measured in eV.

The ratios of the heat fluxes and the electron heat flux, which can be calculated as the coefficients of the thermal and energetic ion heat transport normalized to the coefficient of the electrons, are:
\begin{eqnarray}
\kappa_{ITG,i} = \frac{\langle \Gamma_i \rangle}{\langle \Gamma_e \rangle} = \frac{\alpha_{ITG,i}}{\alpha_{ITG,e}} & = & 2.7   \\
\kappa_{ITG,EP} =\frac{\langle \Gamma_{EP} \rangle}{\langle \Gamma_e \rangle} = \frac{\alpha_{ITG,EP}}{\alpha_{ITG,e}} & = & 0.14
\end{eqnarray}

Although the values of these ratios for the thermal ions and EPs slightly depend on the electron mass, nevertheless they are found to always be respectively higher and lower than unity. This characterises the ITG in the regime of interest.


\section{Dynamics of AMs without turbulence: heat flux}
\label{sec:SAW-heat-flux}

\begin{figure}[b!]
\begin{center}
\includegraphics[width=0.45\textwidth]{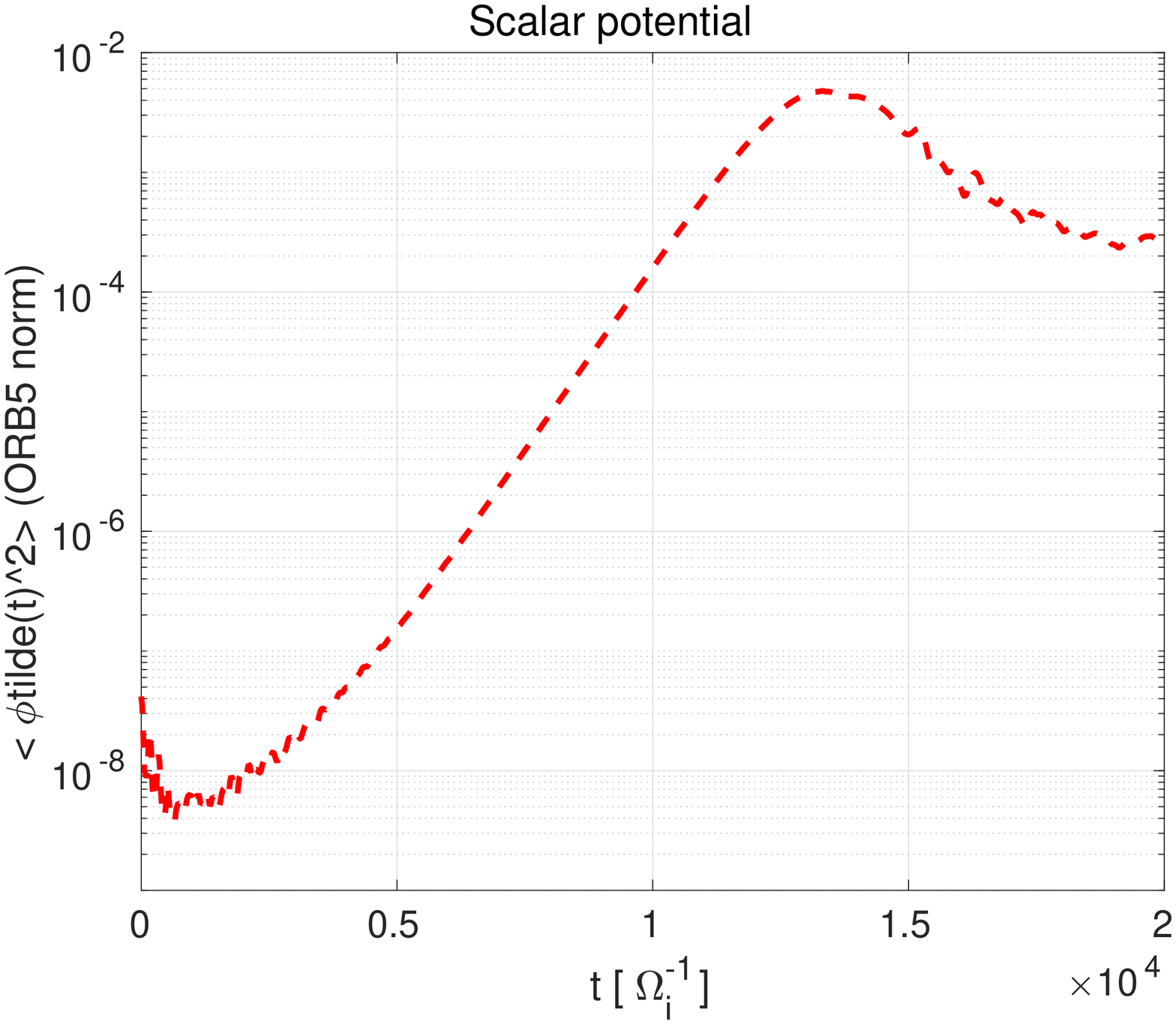}
\includegraphics[width=0.47\textwidth]{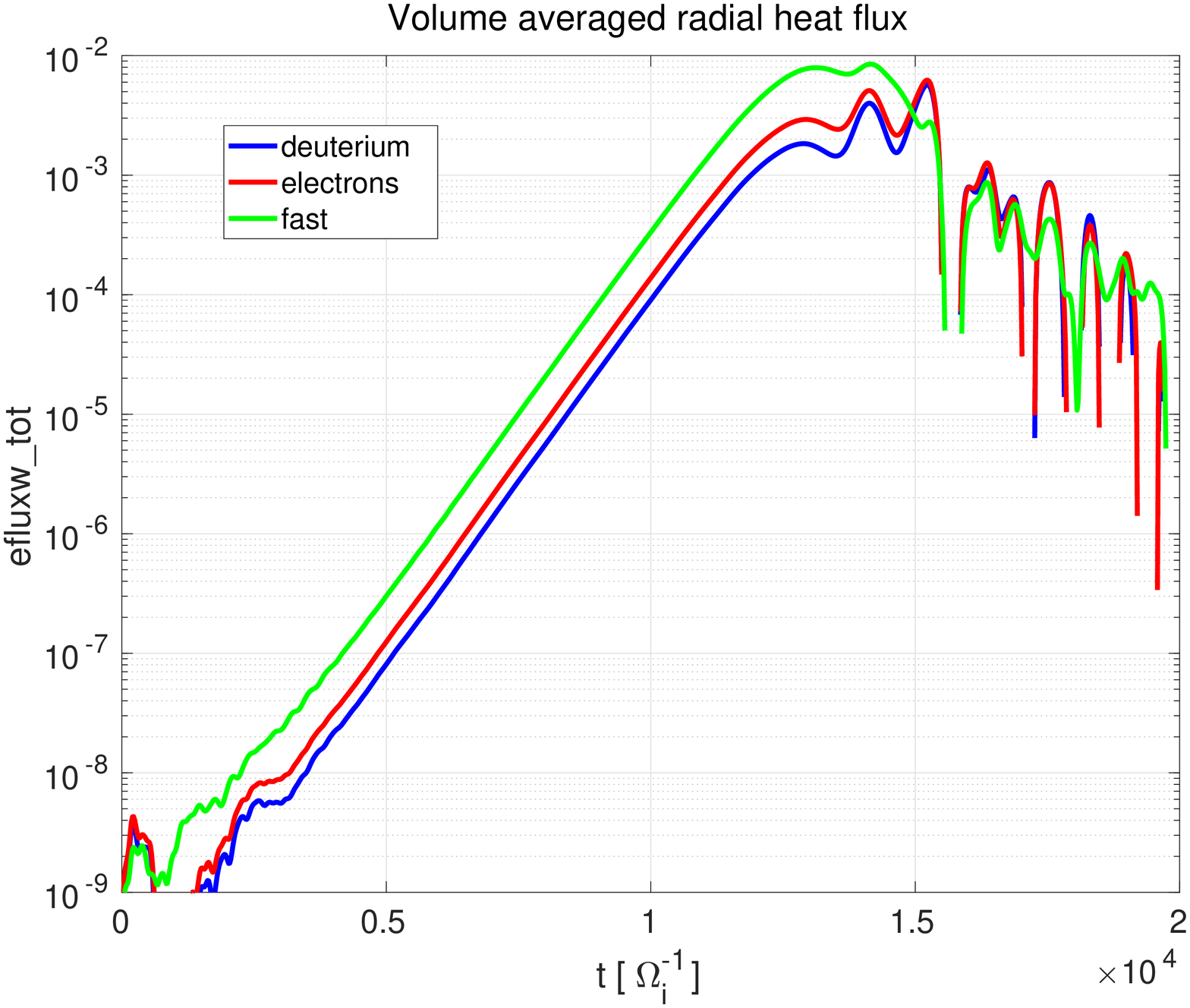}
\includegraphics[width=0.47\textwidth]{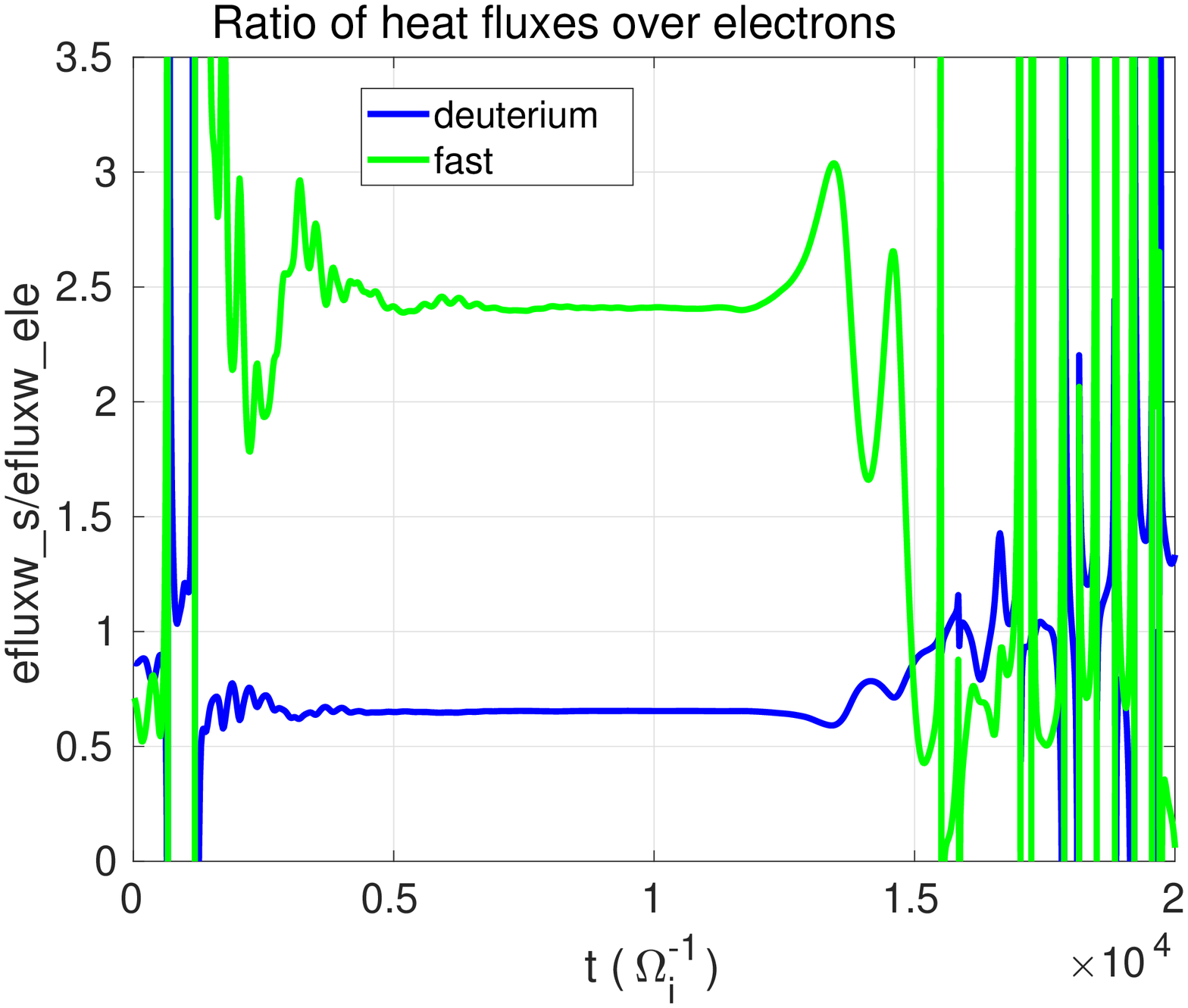}
\caption{\label{fig:SAW-NL-heatflux} Evolution in time of the average of the squared scalar potential in the poloidal plane (top-left), heat fluxes (top-right), and ratios of the heat fluxes of the ions and electrons (bottom). This is a simulation with only $n=5$, fully nonlinear, and with $n_{EP}/n_e=0.01$, representing a BAE.}
\end{center} 
\end{figure}

As an example, we consider here a fully nonlinear  simulation with only $n=5$. No other mode is allowed to develop (therefore there is no turbulence).
The EP concentration is $n_{EP}/n_e=0.01$. The mode is found to grow linearly, saturate at around $t=1.3e4 \; \Omega_i^{-1}$, and damp due to the EP radial redistribution (see Ref.~\cite{Biancalani20JPP} for a detailed analysis of the BAE dynamics in this equilibrium). The evolution in time of the scalar potential and of the heat flux are shown respectively in Fig.~\ref{fig:SAW-NL-heatflux}-top-left and Fig.~\ref{fig:SAW-NL-heatflux}-top-right.

We write the volume averaged heat flux as in Eq.~\ref{eq:heat-flux-ITG-lin}.
In the linear phase, we measure for this BAE with $n=5$ the following values respectively for the thermal ions, electrons, and EPs:
\begin{eqnarray}
\alpha_{BAE,i} & = & 0.023\\
\alpha_{BAE,e} & = & 0.035\\
\alpha_{BAE,EP} & = & 0.083
\end{eqnarray}
Note that, with respect to the ITG (see Sec.~\ref{sec:ITG-lin}), a BAE drives a lower ion heat flux, a similar electron heat flux, and a much higher EP heat flux (1 order of magnitude higher).

The ratios of the heat fluxes and the electron heat flux, which can be calculated as the coefficients of the thermal and energetic ion heat transport normalized to the coefficient of the electrons, are:
\begin{eqnarray}
\kappa_{BAE,i}= \frac{\langle \Gamma_i \rangle}{\langle \Gamma_e \rangle} = \frac{\alpha_{BAE,i}}{\alpha_{BAE,e}} & = & 0.7 \label{eq:kappa_BAE_i}\\
\kappa_{BAE,EP}= \frac{\langle \Gamma_{EP} \rangle}{\langle \Gamma_e \rangle} = \frac{\alpha_{BAE,EP}}{\alpha_{BAE,e}} & = & 2.4 \label{eq:kappa_BAE_EP}
\end{eqnarray}
These characteristic values are found to be stable during the linear phase, and they change after the nonlinear saturation (see Fig.~\ref{fig:SAW-NL-heatflux}-bottom).
They are found to be at convergence with the electron mass for $m_i/m_e=200$.
Note that the EP species carries a heat flux of the same order of magnitude of the thermal species, in these simulations. In order to understand this, we can calculate the value of the ion diamagnetic frequency, for a mode with $n$=5 sitting at s=0.4: $\omega_{*}/\Omega_i = \rho^{*2} nq \kappa_{ni} = 8.8\cdot 10^{-5}$.
The diamagnetic frequency is about two orders of magnitude lower than the mode frequency~\cite{Biancalani20JPP}. So, for our case, we have that an energetic species with the same concentration of the thermal species, is expected to carry a radial flux of about 2 orders of magnitude higher than the thermal species, due to the higher temperature (1 order of magnitude) and higher density gradient (1 order of magnitude). The EP species considered here has a concentration of 2 orders of magnitude lower than the thermal species, which explains the fact that the heat flux is of the same order of magnitude of the thermal heat flux.

The reason why a BAE can carry a significant amount of heat flux of the thermal species, with respect to other AMs, is due to its relatively low frequency, which increases the importance of the wave-particle resonances~\cite{Zonca96}. Regarding the thermal ions, the dominant resonance is expected to be with the transit frequency. Regarding the electrons, which typically have a much higher thermal velocity, the resonances are expected to occur mainly with barely trapped electrons, similarly to what happens for EGAMs~\cite{Novikau19}. The strong electron transport due to AMs found here with ORB5 simulations, is consistent with earlier theoretical predictions~\cite{Briguglio00} (see also Ref.~\cite{Chen99} for a treatment in non-toroidal magnetic equilibria) and experimental observations~\cite{Stutman09}.
%


\section{Nonlinear dynamics of AMs with turbulence}
\label{sec:EM-turb-AMs}

%
%
%

\subsection{Evolution of the fields}

Here, the results of the nonlinear simulations with turbulence and AMs are shown. The restart with EPs switched on is performed at $t\simeq4.9e4 \, \Omega_i^{-1}$.
Before discussing the results of the simulations where the EPs have their nominal temperature, i.e. $T_{EP}(s_r)/T_e(s_r)=10$, we have run a simulation where the EP switch is performed only by increasing the density profile to $\kappa_n=10$, but keeping the EP temperature equal to the thermal species. This test is done to study the effect of the modification in the density profiles alone. No sensible change is observed. Therefore, we can state that the effect of density gradient alone of the minority with thermal temperature, is negligible. 

\begin{figure}[t!]
\begin{center}
\includegraphics[width=0.49\textwidth]{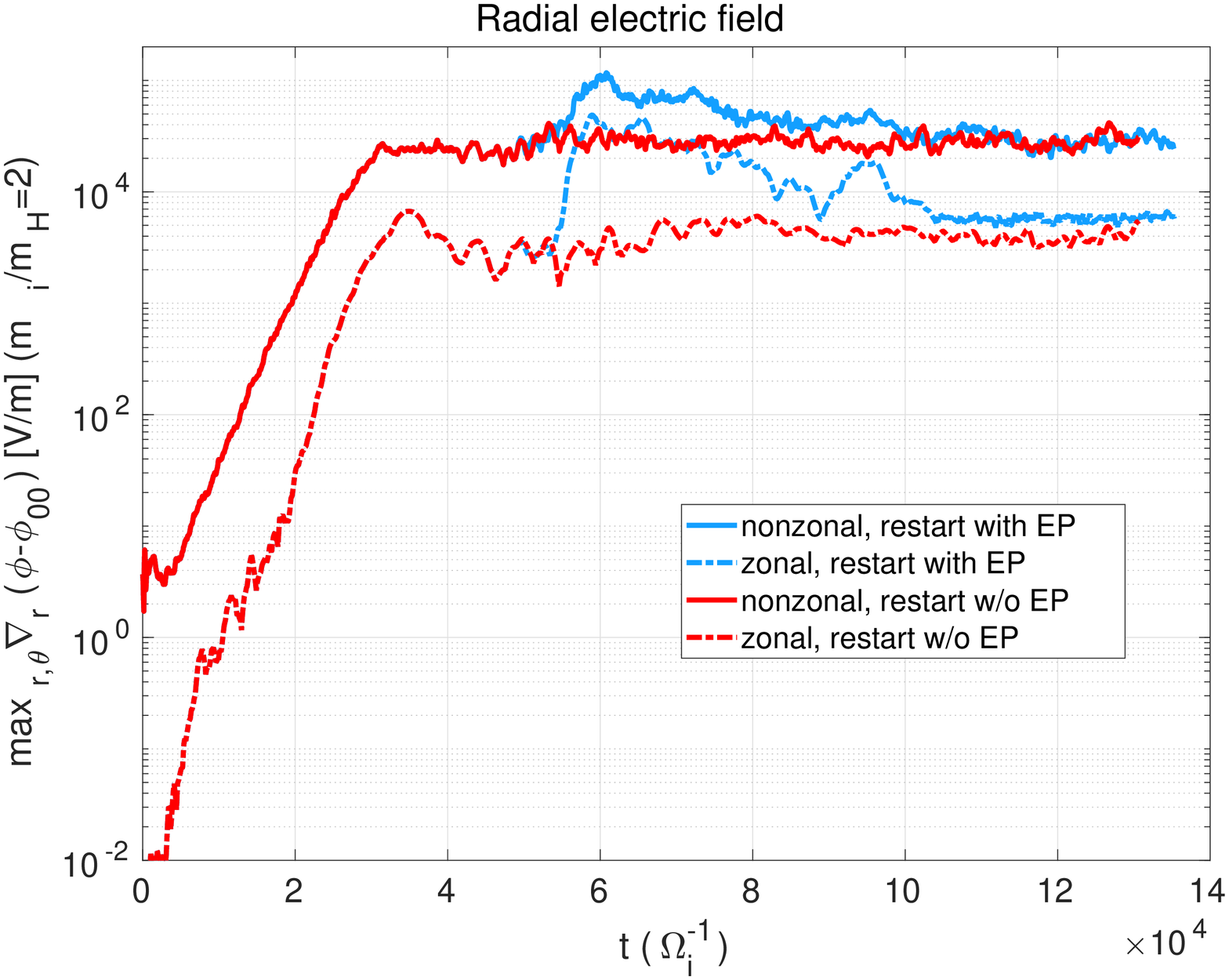}
\includegraphics[width=0.49\textwidth]{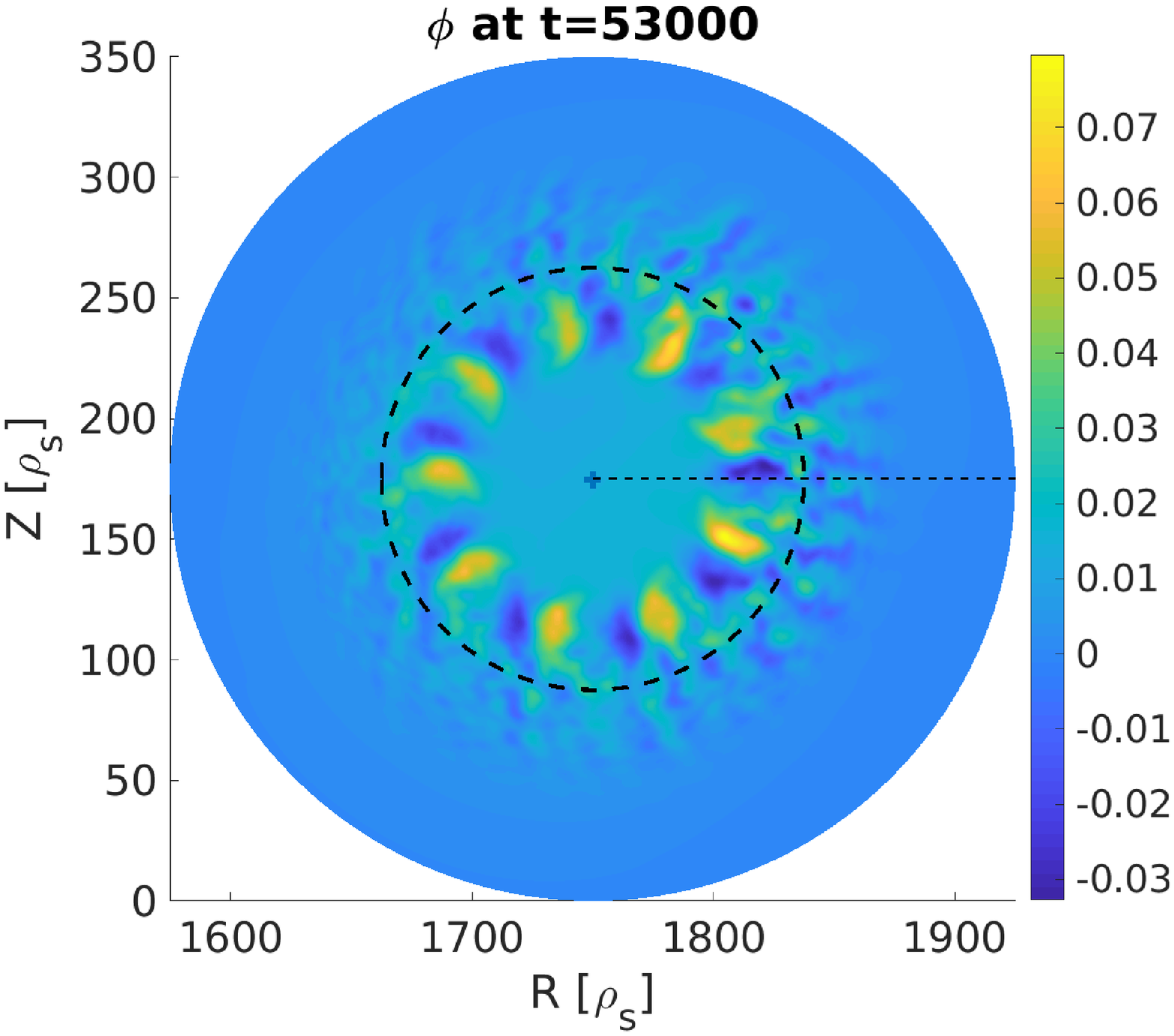}
\vskip -1em
\caption{\label{fig:ITG-SAW-NL-comparison-with-wo-EP} Evolution in time of the maximum of the radial electric field in the poloidal plane (left). Blue lines depict sims where EPs are swiched on, whereas red lines depict simulations where EPs are not switched on. Scalar potential at t=53000 (right), after switching on the EPs.}
\end{center}
\end{figure}

We can now study the effect of the EPs with nominal temperature i.e. $T_{EP}(s_r) / T_e(s_r) = 10$.
In Fig.~\ref{fig:ITG-SAW-NL-comparison-with-wo-EP}-a, the evolution in time of the maximum of the radial electric field is shown, for a simulation where the EPs are switched on at the restart, and compared with a simulation where the EP are not switched on at the restart. One can see that the effect of the EPs is to excite both the nonzonal and zonal components of the electric field, which start growing, then saturate around  $t\simeq 6e4 \, \Omega_i^{-1}$, and then damp in time.

The radial structure shows that the BAE mode with n=5, m=9 is excited by the EPs in the second part of the simulation, when the EPs are switched on (see Fig.~\ref{fig:ITG-SAW-NL-comparison-with-wo-EP}-b).
This BAE is not clearly visible before the EPs are switched on. This mode is identified as a BAE due to the polarization, i.e. a clear n=5, m=9 signature, and the frequency, which is slightly higher than the linear BAE frequency, i.e. two orders of magnitude higher than the frequency of the ITG with n=5 (see Sec.~\ref{sec:ITG-lin}). This frequency is observed at each radius in the domain where the linear BAE and the linear ITG are observed.  Therefore, this BAE is shown to be dominant in amplitude on the ITG modes in this time window (see Fig.~\ref{fig:ITG-SAW-NL-comparison-with-wo-EP}-a). The saturation level of the BAE for this value of krook is $E_r\simeq 1.0\cdot 10^5 \; V/m$ (lowering the value of the krook corresponds to decreasing the artificial damping and reaching a higher saturation level). This value is found to be the same as in the absence of turbulence (see Ref.~\cite{Biancalani20JPP}).

\subsection{Evolution of the total heat fluxes}

\begin{figure}[t!]
\begin{center}
\includegraphics[width=0.49\textwidth]{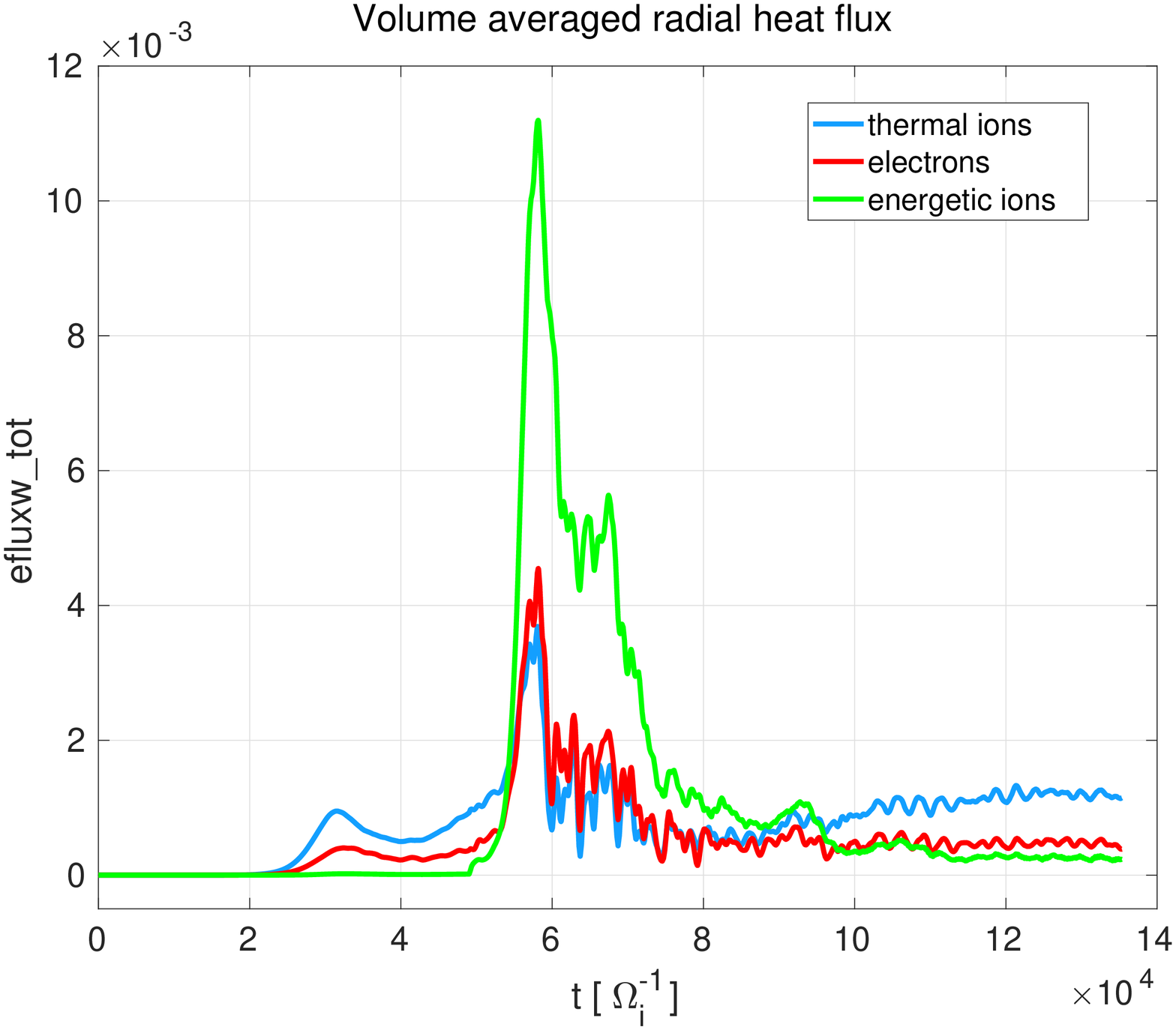}
\includegraphics[width=0.49\textwidth]{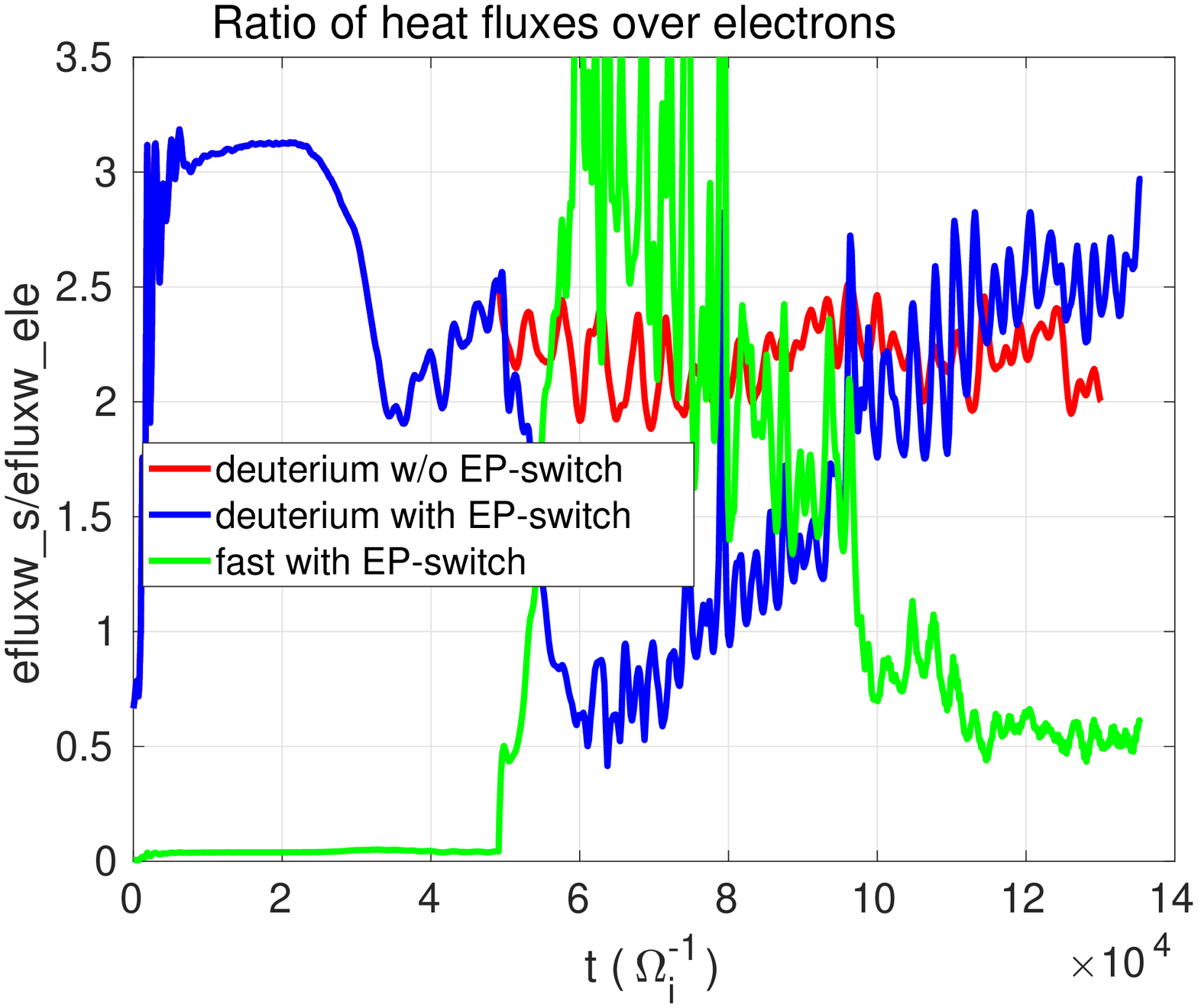}
\vskip -1em
\caption{\label{fig:ITG-SAW-NL-eflux} Heat fluxes of the different species for the simulation of turbulence, with restart with EP (left). The Krook operator is applied here to the thermal species. On the right, the ratio of the heat fluxes of the thermal ions and energetic ions divided by the heat flux of the electrons is shown, for a simulation of AMs and turbulence (and for a simulation w/o EPs, in red, for comparison).}
\end{center} 
\end{figure}

The volume averaged radial heat flux can be studied in a simulation with EPs (see Fig.~\ref{fig:ITG-SAW-NL-eflux}-left). 
At the moment of the BAE saturation, very similar levels of the heat fluxes of the turbulence simulations are obtained with the simulations which retains only $0\le n \le 10$.
If we filter out the $n=10$ mode, slightly lower values of the thermal heat fluxes are obtained. This means that these heat fluxes are not due to the high-n ITG modes, but to the $n=5$ and $n=10$ modes.

The ratios of the heat fluxes with the electron heat flux,
are given in Fig.~\ref{fig:ITG-SAW-NL-eflux}-right. We note that the heat fluxes go from an ITG dominated regime, to a BAE dominated regime, and then back to ITG. In particular, in the first time range when the EPs are not yet switched on, we have $\Gamma_i/\Gamma_e$ bigger than 1 ($\Gamma_i/\Gamma_e>\simeq 3$), which goes down to values lower than unity when the EPs are switched on and the heat flux is dominated by the BAE ($\Gamma_i/\Gamma_e>\simeq 0.7$), and then back to values higher than unity when the BAE is damped. Viceversa, $\Gamma_{EP}/\Gamma_e>$ starts with values higher than unity when they are switched on, indicating a dominant Alfv\'enic activity, to decrease to vaues lower than unity after the BAE has faded away.
%
When sending simulations with $0\le n \le 10$ only, we also try to switch the Krook on, like in the simulation with turbulence, but the values are not sensibly changed. If we consider the simulation where also the mode with $n=10$ is filtered out, the values are not sensibly changed.

\subsection{Evolution of the temperature profiles}

\begin{figure}[t!]
\begin{center}
\includegraphics[width=0.44\textwidth]{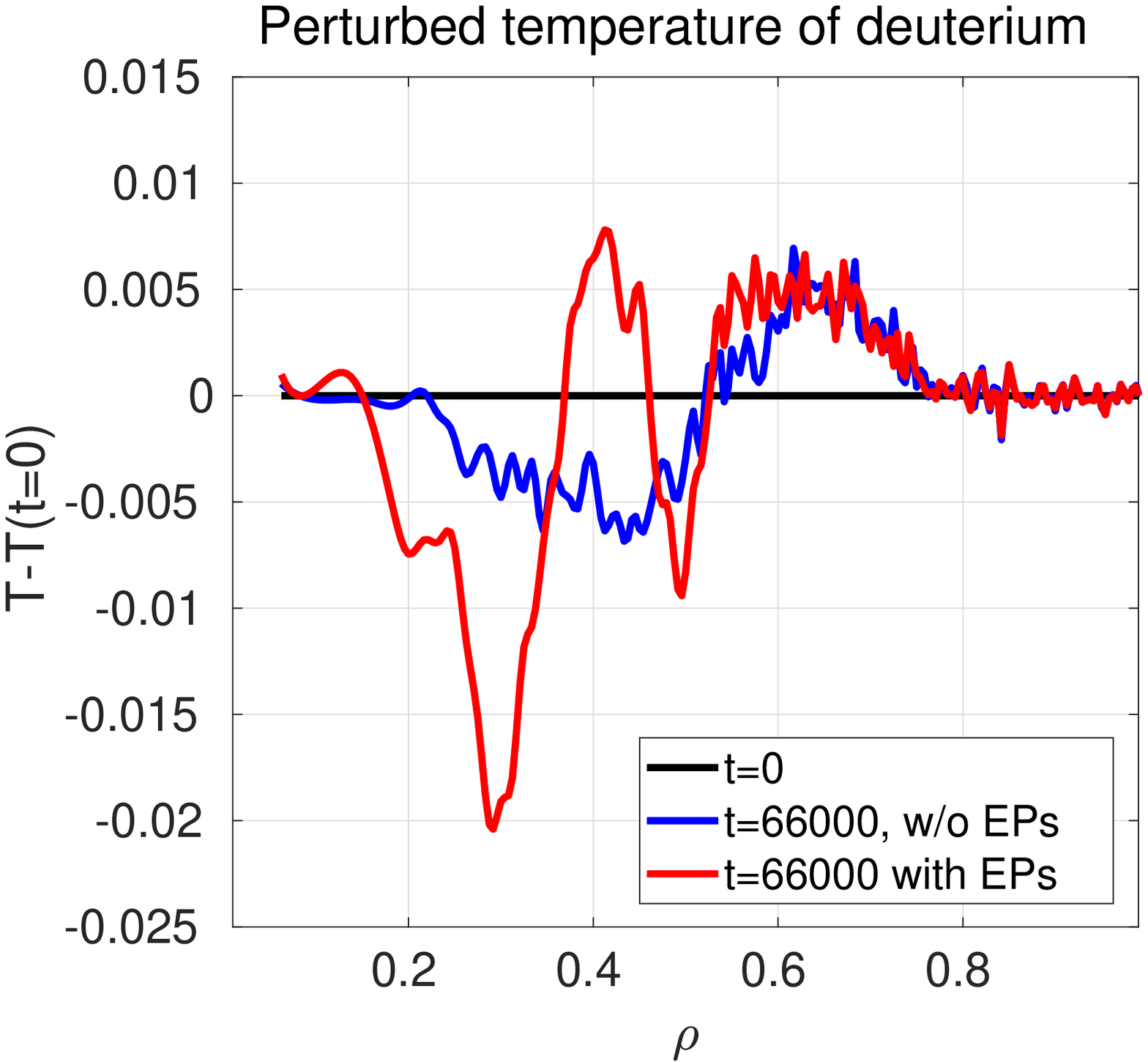}
\includegraphics[width=0.44\textwidth]{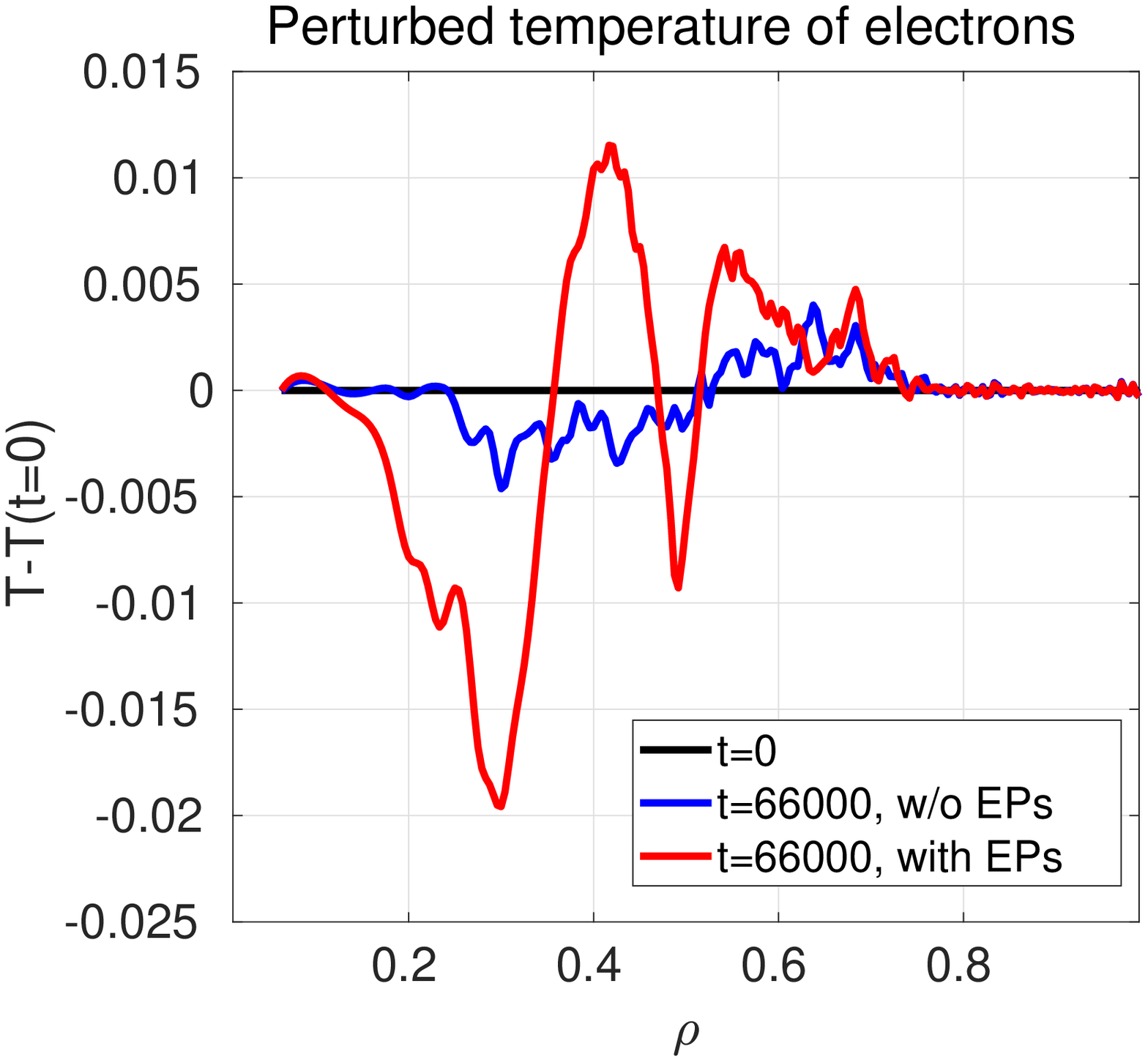}
\vskip -1em
\caption{\label{fig:ITG-SAW-NL-dTprofs} Modification of the temperature profiles of thermal ions (left) and electrons (right) in the simulations with and without EPs, before and after the EPs are switched on.}
\end{center} 
\end{figure}

\begin{figure}[t!]
\begin{center}
\includegraphics[width=0.44\textwidth]{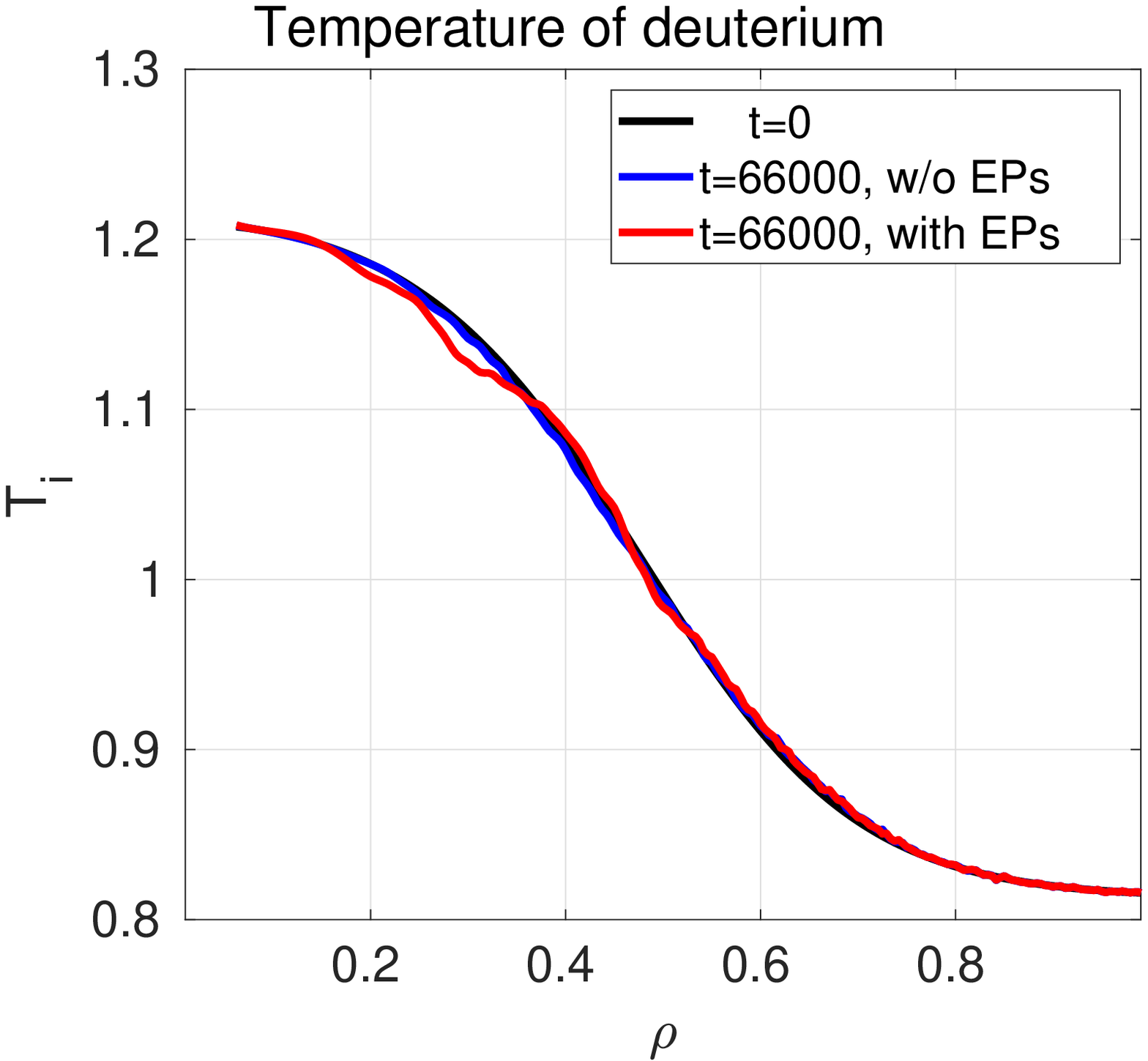}
\includegraphics[width=0.44\textwidth]{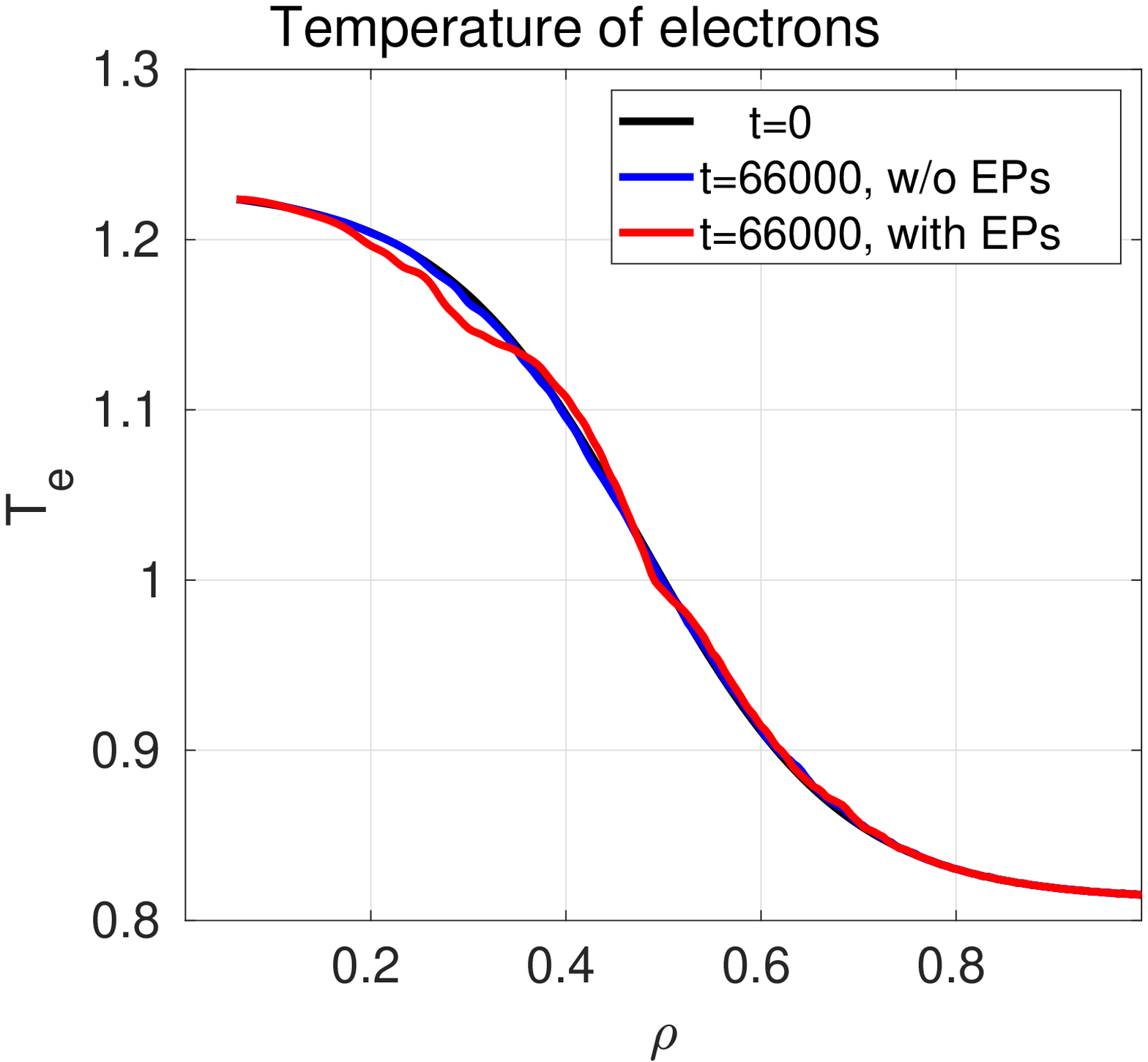}
\vskip -1em
\caption{\label{fig:ITG-SAW-NL-Tprofs-me200-K4} Temperature profiles of thermal ions (left) and electrons (right) in the simulations with and without EPs, before and after the EPs are switched on (with sources). Note the temperature flattening at the location of the BAE.}
\end{center} 
\end{figure}
In this subsection, we show the temperature profile evolution in time in the simulations with and without EPs. The simulation is the same as in the previous section. Sources are applied to the thermal species at the restart in the form of a Krook operator, like in the previous sections.
In Fig.~\ref{fig:ITG-SAW-NL-dTprofs}, the difference of the temperature at a certain time and the temperature at the beginning of the simulation is shown. The original profiles are normalized to their value at the reference radius $s_r=0.525$. The time is chosen around the peak of the BAE amplitude. Around this time, the temperature profile is found to have small oscillations and small radial structures are also formed.

Note that the ITG turbulence redistributes the temperature of about a factor 2 higher for the thermal ions than for the electrons, whereas the BAE redistributes both ions and electrons of about the same amount.
Note also that the amplitude of the ion temperature perturbation of the BAE is about 3 times higher than the ITG.

In Fig.~\ref{fig:ITG-SAW-NL-Tprofs-me200-K4}, the evolution of the temperature profiles is shown. Note the flattening of the profiles around the location of the BAE, due to the heat flux carried by the BAE.

\subsection{Spectra in toroidal mode number}

Both ITG turbulence and AMs are known to induce cross-field heat transport. In the simulations presented here, both of these effects are co-existing, raising two general questions: What is the relative importance of each of these two mechanisms, and do they interfere in any way?

\begin{figure}[h!]
\begin{center}
\includegraphics[width=0.5\textwidth]{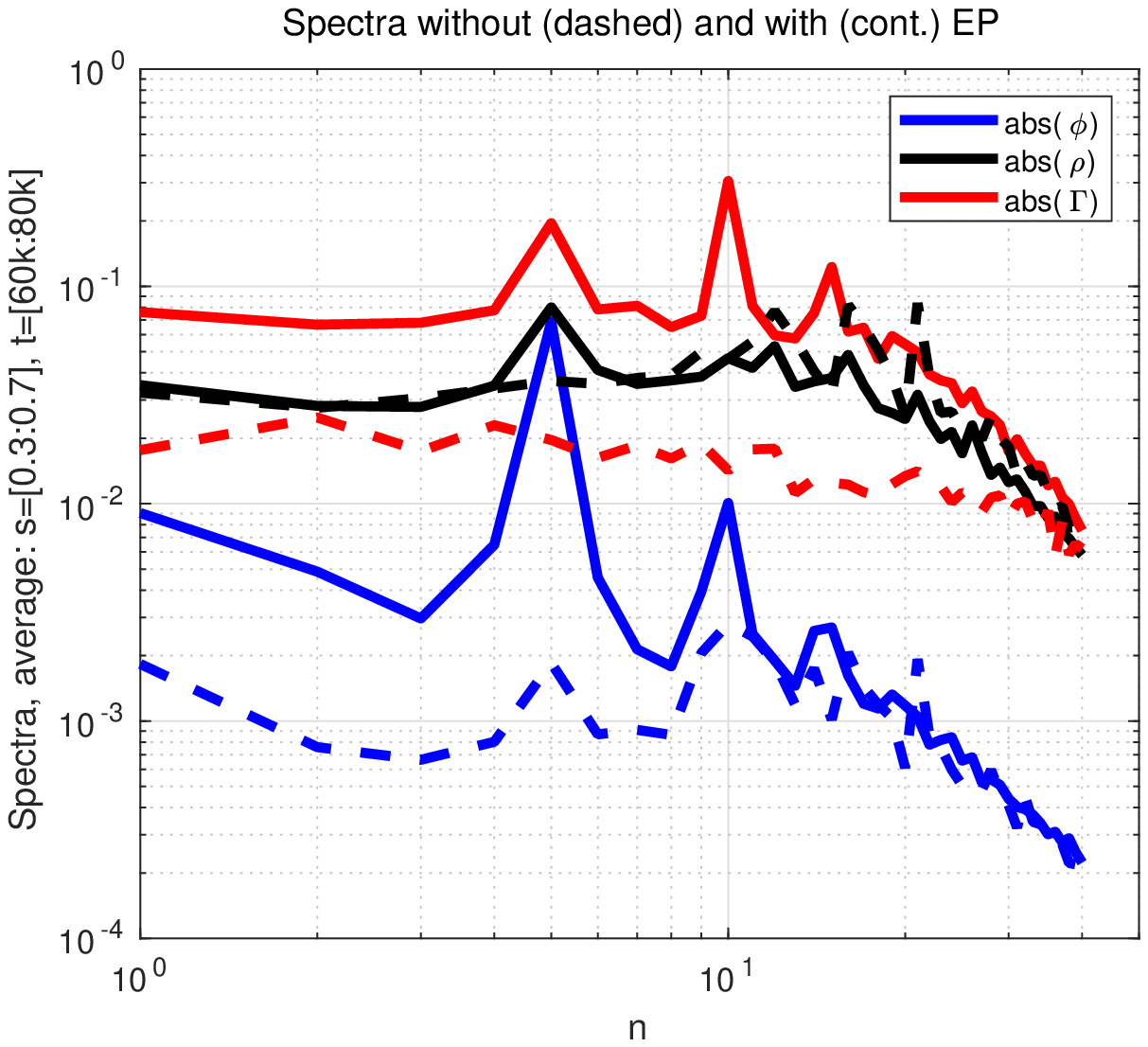}
\includegraphics[width=0.47\textwidth]{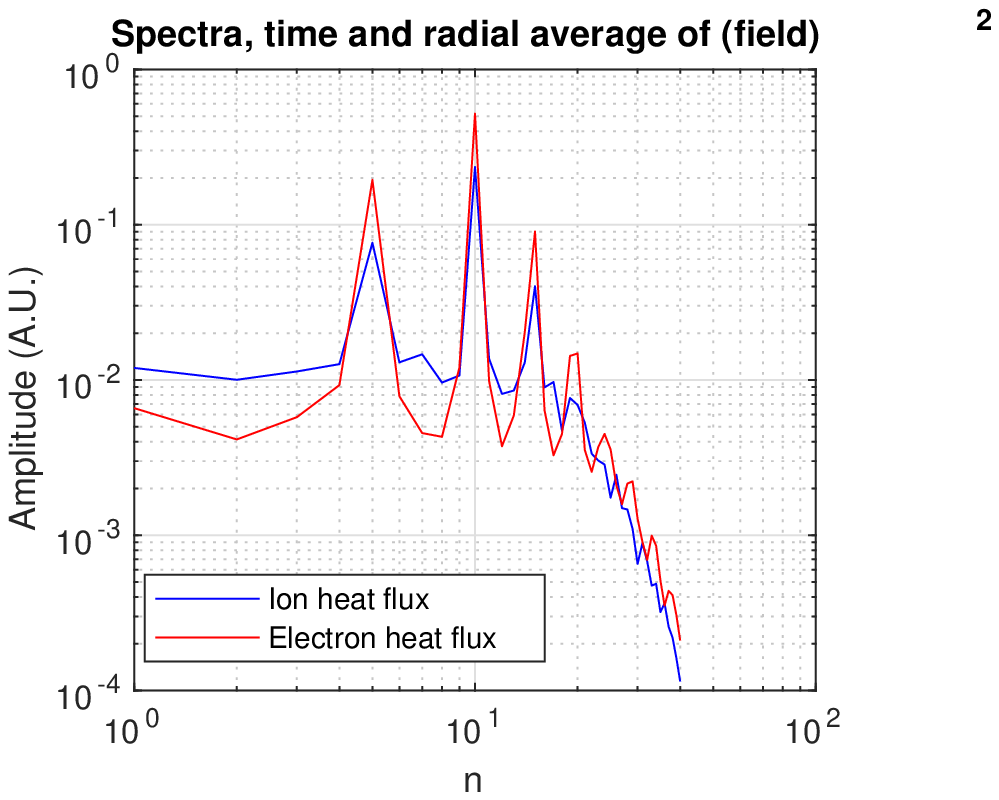}
\vskip -1em
\caption{\label{fig:spectrum_n} 
Toroidal mode number spectra of scalar potential $\phi$, density $\rho$, and ion heat flux $\Gamma$, with and w/o EPs (left). Toroidal mode number spectra of the ion and electron heat flux with EPs (right).}
\end{center}
\vskip -2em
\end{figure}

The toroidal mode number spectrum of the ion heat flux (for the details of the diagnostic, see Ref.~\cite{Hayward19})
for a simulation where the EPs are switched on can be seen as a the continuous red line in Fig.~\ref{fig:spectrum_n}. For comparison, the same spectrum for a simulation without EPs is shown as a dashed red line. A sensible general trend that we find is that the heat fluxes are higher in the simulation with EPs. This is explained by the fact that the EP density profile consitutes an additional source of free energy to the system.

\subsubsection{Analysis of the spectrum at low-$n$}

\begin{figure}[b!]
\begin{center}
\includegraphics[width=0.44\textwidth]{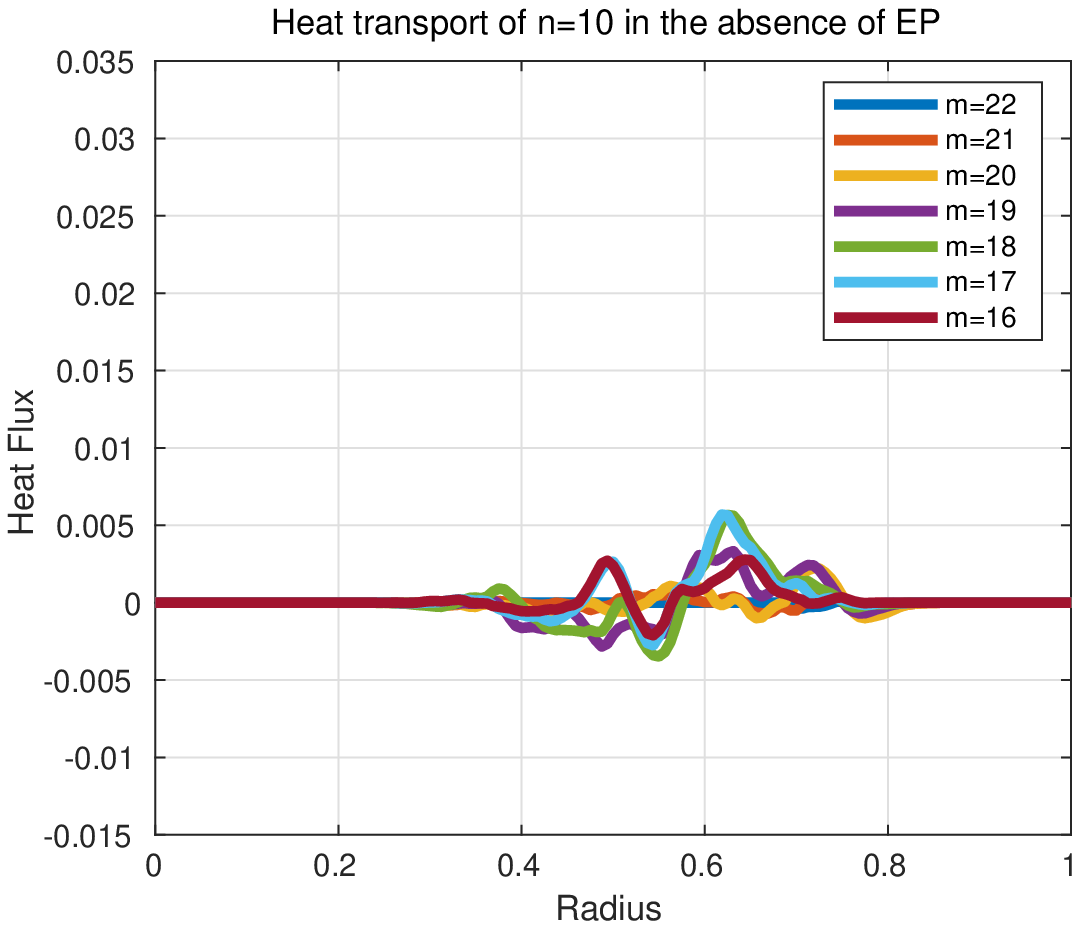}
\includegraphics[width=0.43\textwidth]{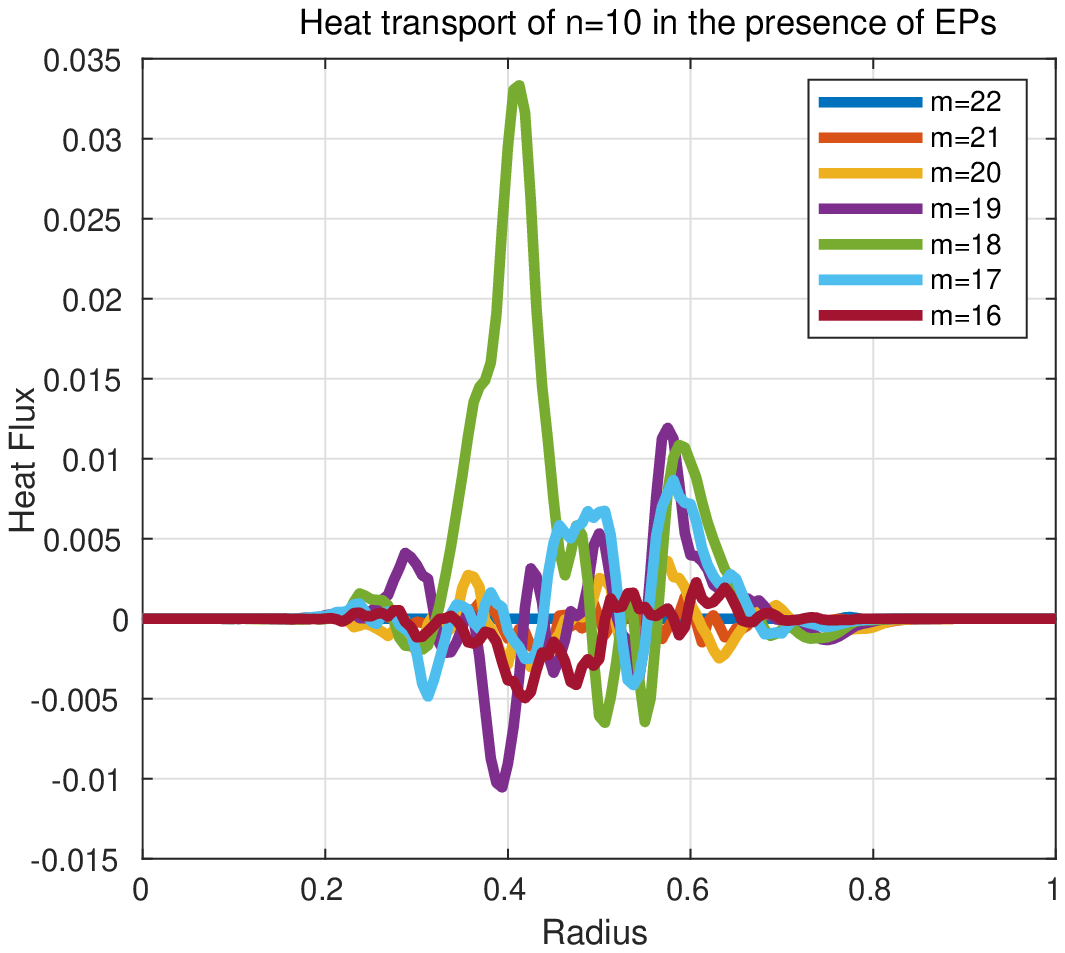}
\vskip -1em
\caption{\label{fig:n10woEP}
Radial structure of heat flux for $n=10$, and for different poloidal modes, for a simulation without (left) and with (right) EPs. Without EPs, the polarization with equal contribution of the different poloidal components identifies the ITG dynamics.
With EPs, the polarization with dominant $m=18$ identifies the BAE second harmonics, on top of the ITG turbulence.
}
\end{center}
\end{figure}

We observe that the nonzonal spectra given in Fig.~\ref{fig:spectrum_n} are dominated by the modes $n=5$, $n=10$ and $n=15$, namely the first, second and third harmonics of the main BAE mode.
The contributions of the different poloidal mode numbers to the heat fluxes for the dominant mode, namely the mode with $n=10$, is depicted for a simulation without and with EPs, in Fig.~\ref{fig:n10woEP}.
In the case of the simulation without EPs, a broad spectrum of poloidal components shows the characteristic polarization of the ITG mode.
On the other hand, for the case of the simulation with EPs, the heat flux is dominated by the poloidal mode $m=18$.  Note also that the peak of the $m=18$ component is at the position of the peak of the BAE field. These two are signatures of the second harmonics of the BAE. Therefore, we can state that the ITG dynamics is subdominant in the simulation with EPs, with respect to the BAE, in carrying the heat transport.

It is also important to note that the other poloidal components (i.e., $m\ne18$) are increased in the simulation with EPs, in comparison with the simulation without EPs. This means that the BAE second harmonic is efficient in modifying the dynamics of the ITG of $n=10$, due to the nonlinear interaction. This is an example of cross-scale interaction, with the AM being the macro-scale mode, and the ITG turbulence being constituted mainly by micro-instabilities.

\subsubsection{Analysis of the spectrum at high-$n$}

\begin{figure}[b!]
\begin{center}
\includegraphics[width=0.44\textwidth]{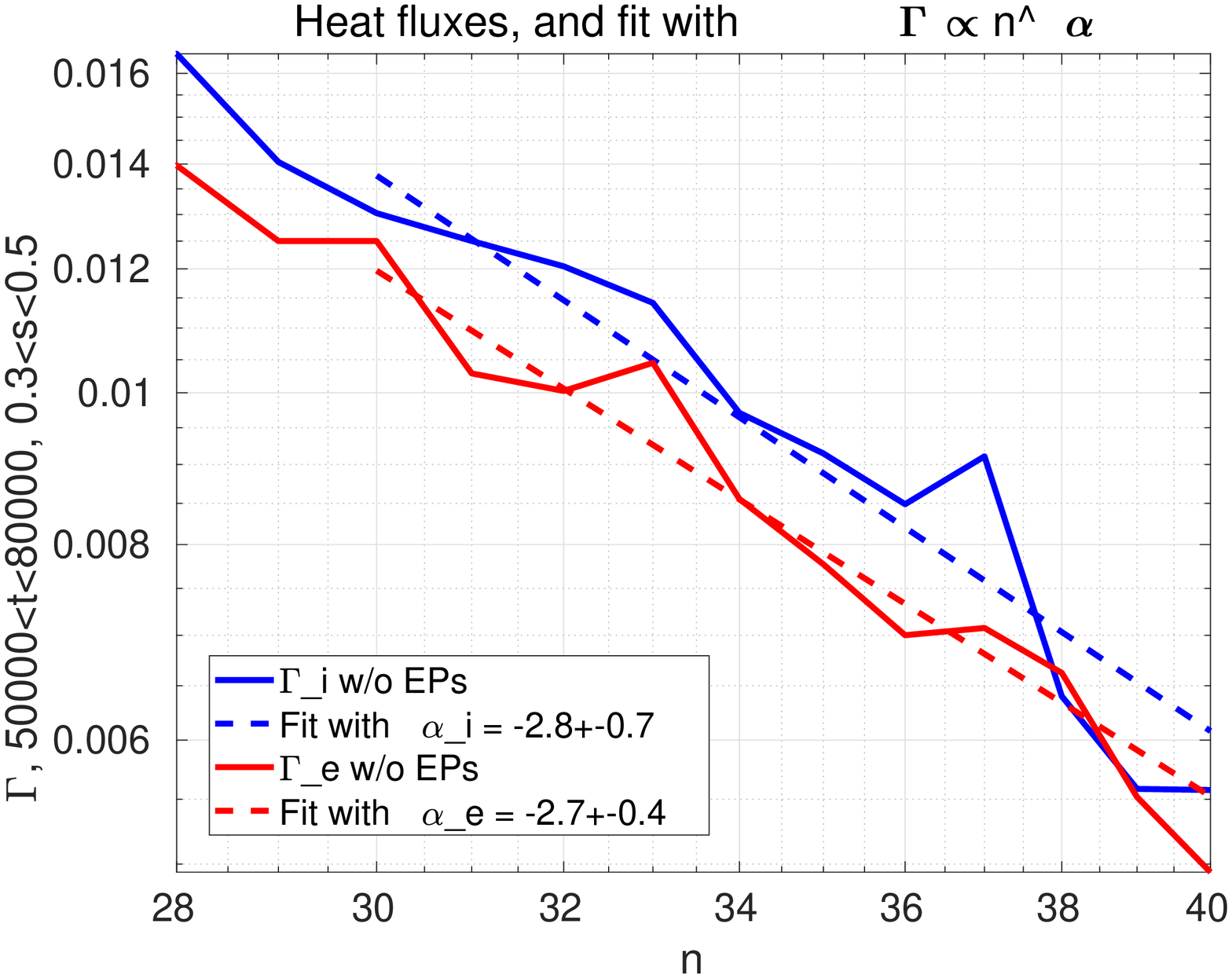}
\includegraphics[width=0.44\textwidth]{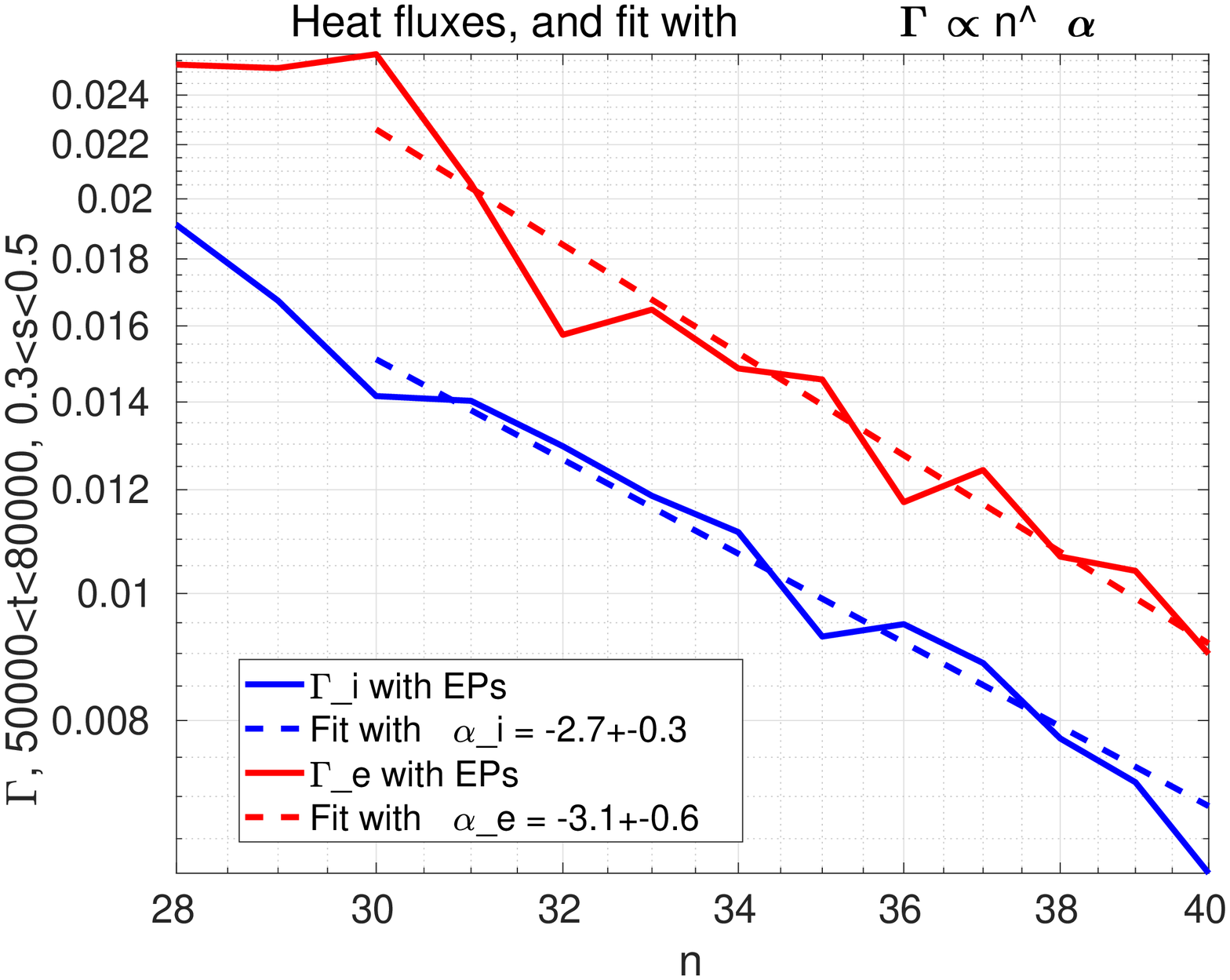}
\vskip -1em
\caption{\label{fig:Gammaie-me200-K4-a}
Heat fluxes of ions (blue) and electrons (red) in the simulation w/o EPs (left) and with EPs (right), with $m_i/m_e$=200, with sources. The measurement is done around the location of the BAE, i.e. in the region $0.3<s<0.5$. Note the effect of the EPs in reversing the importance of the electron over the ion heat fluxes.
}
\end{center}
\end{figure}

\begin{figure}[t!]
\begin{center}
\includegraphics[width=0.44\textwidth]{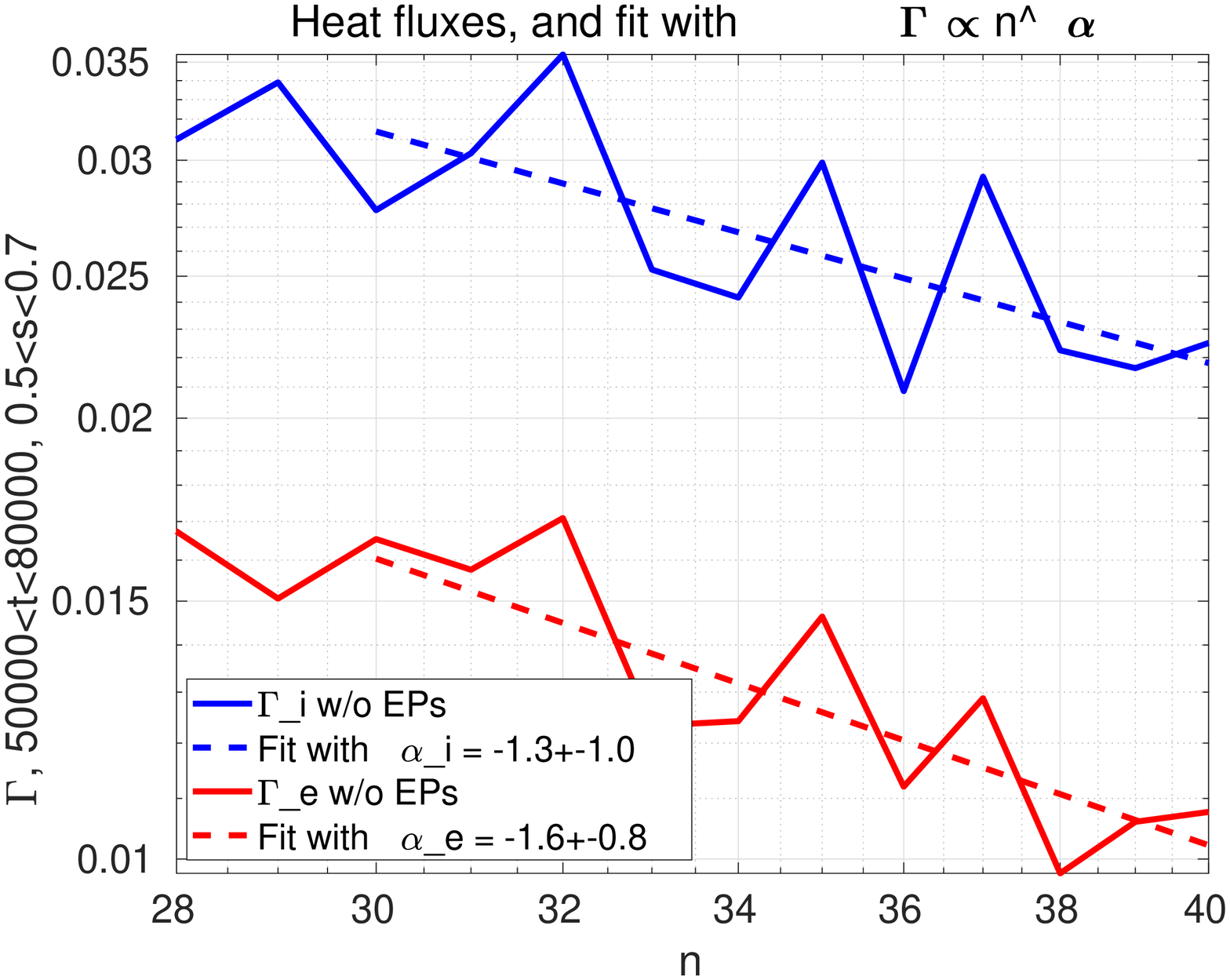}
\includegraphics[width=0.44\textwidth]{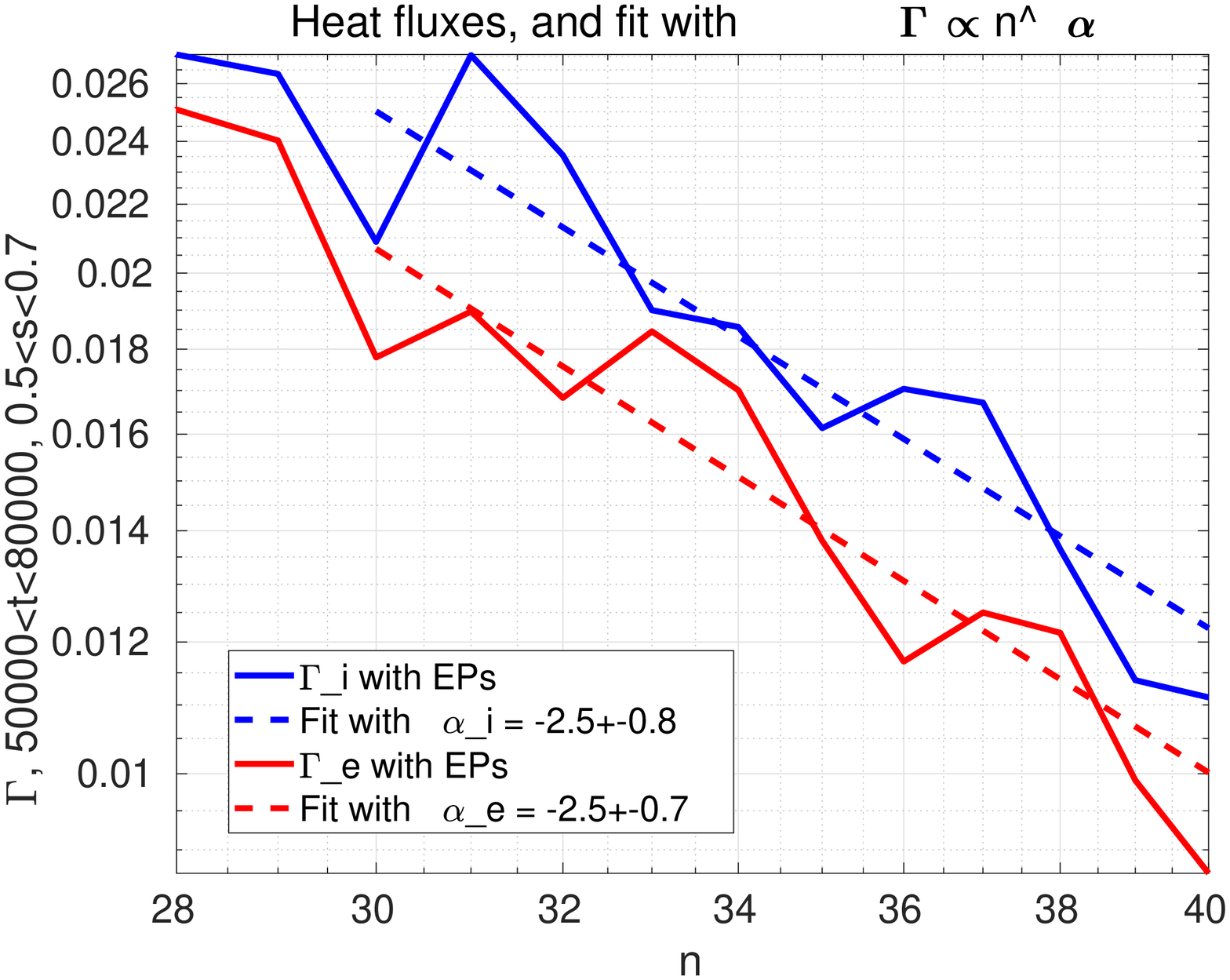}
\vskip -1em
\caption{\label{fig:Gammaie-me200-K4-b}
Same as in Fig.~\ref{fig:Gammaie-me200-K4-a}, but with a measurement done far from the location of the BAE, namely in the region $0.5<s<0.7$. Here the ion heat flux is always bigger than the electron heat flux, as the ITG turbulence dominates the spectra.
}
\end{center} 
\end{figure}

It is worthy to note that not only the low-$n$ part of the heat flux spectrum, i.e., where the BAEs are dominant, has higher levels in the presence of EPs, but also the higher-$n$ part, i.e., where the ITGs are dominant. This is of interest for the question of how the presence of EPs modifies the ITG turbulence. In this regime, we can state that an EP population driving a BAE linearly unstable, affects the turbulence dynamics by increasing the heat transport.
The fluctuation amplitude of the scalar potential $\phi$ and density $\rho$ for the different toroidal mode numbers is also shown in Fig.~\ref{fig:spectrum_n} (respectively with blue lines and blue black lines). Differently from the heat flux, the fluctuation amplitude given by the scalar potential is shown not to be sensibly modified by the presence of EPs in the range of toroidal mode numbers of the ITG. Moreover, the fluctuation amplitude given by the perturbed density is found to be decreased by the presence of EPs. 

We can also investigate the relative importance of the ion and electron heat fluxes in the domain of high-$n$. This is shown in Fig.~\ref{fig:Gammaie-me200-K4-a} and \ref{fig:Gammaie-me200-K4-b}. We note that the ion heat flux is always bigger than the electron heat flux in the radial region $0.5<s<0.7$, with or without EPs, because the spectrum is dominated by the ITG turbulence. On the other hand, the electron heat flux is bigger than the ion heat flux in the radial region $0.3<s<0.5$, when the EPs are switched on, because in this region the spectrum becomes dominated by the BAE.


\section{Analytical estimation of the heat fluxes of ballooning and non-ballooning modes}
\label{sec:anal}

In the previous sections, we have described the results of global gyrokinetic simulations of AMs and ITG turbulence. We have shown that AMs drive a strong electron heat flux, in comparison with ITGs, which have the ion heat flux dominant. AMs and ITGs differ mainly because of their frequency, spatial structure, and polarization.

In this Section, we want to investigate how the spatial structure of the modes influences the electron heat flux, by estimating the heat flux analytically. To this aim, we consider two types of modes. The first is an ITG: a mode with ballooning structure, i.e. with higher amplitude on the low-field side of the tokamak and lower amplitude at the high-field side of the tokamak. The second is a BAE: a mode with non-ballooning structure, i.e. with equivalent intensity on the low-field side and high-field side.

Let us recall that the distribution function of each species $s$
is written as 
\begin{equation}
f_{s}=f_{Ms}({\bf r},v)\left(1-\frac{e_{s}\phi({\bf r},t)}{T_{s}}\right)+g_{s}({\bf R},v,\lambda,t), 
\end{equation}
where ${\bf r}$ is the particle position, ${\bf R}$ the guiding-centre
position, $\phi$ the electrostatic potential, $e_{s}$ the charge,
$v$ the speed, $\lambda=v_{\perp}^{2}/(v^{2}B)$ the ratio between
the magnetic moment $\mu=m_{s}v_{\perp}^{2}/(2B)$ and the kinetic
energy $m_{s}v^{2}/2=x^{2}T_{s}$, and 
\[
f_{Ms}=n_{s}\left(\frac{m_{s}}{2\pi T_{s}}\right)^{3/2}e^{-x^{2}}
\]
the Maxwellian with particle density $n_{s}(\psi)$ and temperature
$T_{s}(\psi)$. In this notation, the equation for $g_{a}$ becomes
\begin{equation}
iv_{\|}\nabla_{\|}g_{s}+(\omega-\omega_{da})g_{s}=J_{0}\left(\frac{k_{\perp}v_{\perp}}{\Omega_{s}}\right)\frac{e_{s}\phi}{T_{s}}\left(\omega-\omega_{*s}^{T}\right)f_{Ms},\label{gk}
\end{equation}
where $\Omega_{s}=e_{s}B/m_{s}$ is the gyrofrequency and the derivatives
are taken at fixed energy and magnetic moment. $J_{0}$ is the zeroth-order
Bessel function of the first kind, and corresponds to a gyroaveraging
operator in real space. The magnetic field is taken to be ${\bf B}=\nabla\psi\times\nabla\alpha$,
${\bf k}_{\perp}=k_{\psi}\nabla\psi+k_{\alpha}\nabla\alpha$, where in this section $\psi$ is the toroidal magnetic flux, and $\alpha=q\theta - \varphi$, with $\theta$ and $\varphi$ respectively the poloidal and toroidal angles.
The diamagnetic and drift frequencies, respectively, are defined by
\[
\omega_{\ast s}=\frac{k_{\alpha}T_{s}}{e_{s}}\frac{d\ln n_{s}}{d\psi},
\]
\[
\omega_{\ast s}^{T}=\omega_{\ast s}\left[1+\eta_{s}\left(x^{2}-\frac{3}{2}\right)\right],
\]
\[
\omega_{ds}={\bf k}_{\perp}\cdot{\bf v}_{ds},
\]
where $\eta_{s}=d\ln T_{s}/d\ln n_{s}$ and ${\bf v}_{ds}=\Omega_{s}^{-1}\mathbf{b}\times\left[v_{\parallel}^{2}\mathbf{b}\cdot\nabla\mathbf{b}+0.5v_{\perp}^{2}\nabla B/B\right]$
denotes the drift velocity. 

Within this framework, we introduce the flux-surface averaged
quasilinear radial heat flux, for species $s$: 
\begin{equation}
\begin{split} & \Gamma_{s}=\Re\left\langle \int d^{3}\mathbf{v}\delta f_{s}\frac{m_{s}v^{2}}{2}\mathbf{v}_{E}^{*}\cdot\nabla\psi\right\rangle _{\psi}\\
 & =-k_{\alpha}\Im\left\langle \int d^{3}\mathbf{v}g_{s}\frac{m_{s}v^{2}}{2}J_{0}\phi^{*}\right\rangle _{\psi}.
\end{split}
\label{eq:QL}
\end{equation}
The kinetic function $g_{s}$ will be considered in the frequency
regime 
\begin{equation}
\omega\sim\omega_{ds}\ll\omega_{be},\omega_{tr,e},
\end{equation}
where $\omega_{be},$ and $\omega_{tr,e}$ are the bounce and transit
frequency. This allows us to solve analytically Eq. \eqref{gk} for
the electrons.

In the following, we will focus on the electron transport, which is found to be dominant for BAEs in the simulations.

\subsection{Passing electrons}

For passing electrons the streaming term is dominant, hence~\cite{Zocco15} 
\begin{equation}
g_{e}^{(p)}=-\left(1-\frac{\omega_{*e}^{T}}{\omega}\right)\frac{e\hat\psi}{T_{e}}f_{Me},
\end{equation}
where $\hat\psi$ is so that $A_{\parallel}=c\nabla_{\parallel}\hat\psi/(i\omega).$
By replacing this expression in Eq. (\ref{eq:QL}), after writing
$\omega=\omega_{r}+i\gamma,$ one finds
\begin{equation}
\begin{split} & Q_{e}^{(p)}=\frac{ek_{\alpha}}{T_{e}}\left\langle \int d^{3}\mathbf{v}f_{Me}\frac{m_{e}v^{2}}{2}\left[\left(1-\frac{\omega_{*e}^{T}\omega_{r}}{\left|\omega\right|^{2}}\right)\left(\Im\hat\psi\Re J_{0}\phi-\Re\hat\psi\Im J_{0}\phi\right)\right.+\right.\\
 & \left.\left.\frac{\omega_{*e}^{T}\gamma}{\left|\omega\right|^{2}}\left(\Re\hat\psi\Re J_{0}\phi+\text{\ensuremath{\Im}}\hat\psi\Im J_{0}\phi\right)\right]\right\rangle _{\psi}.
\end{split}
\label{eq:Qepass}
\end{equation}
In the specific case $E_{\parallel}\rightarrow0,$ $\phi=\hat\psi,$ and
for a periodic potential in $\theta,$ e.g. $\phi=\cos\theta+i\sin\theta,$
we have 
\begin{equation}
\begin{split} & Q_{e}^{(p)}=\frac{ek_{\alpha}}{T_{e}}\left\langle \left(\Re\hat\psi\Re J_{0}\phi+\Im\hat\psi\Im J_{0}\phi\right)\frac{\gamma}{\left|\omega\right|^{2}}\int d^{3}\mathbf{v}f_{Me}\frac{m_{e}v^{2}}{2}\omega_{*e}^{T}\right\rangle _{\psi}\\
 & =\frac{3}{2}ek_{\alpha}n_{0}\frac{\gamma\omega_{*e}}{\left|\omega\right|^{2}}(1+\eta_{e}).
\end{split}
\end{equation}
Then, close to marginality, passing particles do not give a contribution
to transport.

\subsection{Trapped electrons}

We consider trapped electrons to be mostly electrostatic.
We bounce average the electron kinetic equation to obtain 
\begin{equation}
\begin{split} & g_{e}^{(tr)}=-\frac{\omega-\omega_{*e}^{T}}{\omega-\bar{\omega}_{de}}\frac{ef_{Me}}{T_{e}}\overline{\phi}\end{split}
\label{eq:soltrap}
\end{equation}
where we took $J_{0}=1,$ since we are interested in transport at
the ion scale, and the bounce-average is defined in the appendix.
By inserting Eq. (\ref{eq:soltrap}) in (\ref{eq:QL}), one obtains
\begin{equation}
\begin{split} & \Gamma_{e}=-n_{e}\left(\chi_{e,n}\frac{1}{n_{e}}\frac{dn_{e}}{d\psi}+\chi_{e,T}\frac{1}{T_{e}}\frac{dT_{e}}{d\psi}+\chi_{e,C}\right)\\
\\
\end{split}
\end{equation}
with 
\begin{equation}
\chi_{e,n}=\frac{\pi k_{\alpha}^{2}}{n_{e}}\left\langle \int_{tr}d^{3}\mathbf{v}f_{Me}\frac{m_{e}v^{2}}{2}\Delta_{\gamma}\left(\omega_{r}-\bar{\omega}_{de}\right)\left|\overline{\phi}\right|^{2}\right\rangle _{\psi},\label{eq:chiN}
\end{equation}
\begin{equation}
\chi_{e,T}=\frac{\pi k_{\alpha}^{2}}{n_{e}}\left\langle \int_{tr}d^{3}\mathbf{v}f_{Me}\frac{m_{e}v^{2}}{2}\left(\hat{v}^{2}-\frac{3}{2}\right)\Delta_{\gamma}\left(\omega_{r}-\bar{\omega}_{de}\right)\left|\overline{\phi}\right|^{2}\right\rangle _{\psi},\label{eq:chiT}
\end{equation}
and 
\begin{equation}
\chi_{e,C}=\frac{\pi ek_{\alpha}}{n_{e}T_{e}}\left\langle \int_{tr}d^{3}\mathbf{v}f_{Me}\frac{m_{e}v^{2}}{2}\bar{\omega}_{de}\Delta_{\gamma}\left(\omega_{r}-\bar{\omega}_{de}\right)\left|\overline{\phi}\right|^{2}\right\rangle _{\psi},\label{eq:chiC}
\end{equation}
We evaluate term by term, using a simple model for the equilibrium
field $B=B_{0}(1-\epsilon\cos\theta),$ in the marginal limit $\gamma\rightarrow0^{+}.$
Results are in the Appendix, and will be used in the following.

\subsection{Influence of the space structure on the electron heat flux}

\begin{figure}[t!]
\begin{center}
\includegraphics[width=0.6\textwidth]{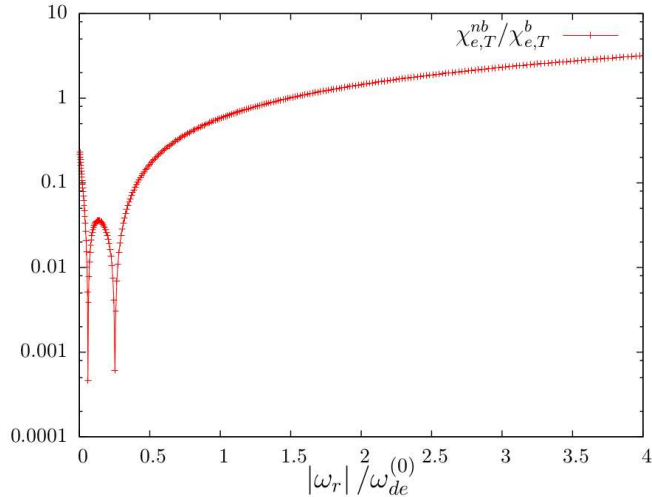}
\vskip -1em
\caption{\label{fig:zocco}
Ratio of the coefficients of the electron heat
flux due to temperature gradients for ITG-like modes and BAE-like
modes. Here $\omega_{r}\omega_{de}^{(0)}<0$.
}
\end{center} 
\end{figure}

We consider two types of modes, one with a maximum on the
outboard-side and a minimum at the inboard-side (ballooning, ITG-like)
\begin{equation}
\phi^{(b)}=\phi_{0}\cos\theta/2,
\end{equation}
one a simple trigonometric function (non-ballooning, BAE-like) 
\begin{equation}
\phi^{(nb)}=\phi_{0}\cos\theta.
\end{equation}

The result is the estimation of the electron heat flux of ballooning and non-ballooning modes as a dependence on the frequency. As an example, we consider the heat flux given by the temperature
gradient term, $\chi_{e,T}$ (the others behave similarly). The ratio of the coefficients of the contributions of non-ballooning and ballooning modes is shown in Fig.~\ref{fig:zocco}. One can see that, below a certain frequency given by $\omega_{r}/\omega_{de}(0)=1.5$, the electron heat flux is dominated by ballooning modes (``b'') like ITGs, whereas above this frequency, the electron heat flux is dominated by non-ballooning modes (``nb'') like BAEs. This shows how the spatial structure of BAEs is important, together with the frequency, in driving an important electron heat flux, in comparison with ITGs.


\section{Conclusions and discussion}
\label{sec:conclusions}

The transport of heat and particles is one of the most important problems for the confinement of tokamak plasmas in present day tokamaks and future devices. Historically, the radial transport of energetic particles (EPs) by Alfv\'en modes (AMs) and the heat transport of the thermal species by turbulence, have been treated separately. Nowadays, it is becoming clearer that these two problems can actually be connected, because their mutual influence can be strong in some regimes. Moreover, it is becoming more feasible to study their selfconsistent interaction by means of numerical simulations, due to the more powerful supercomputers, and to more efficient numerical schemes.
A kinetic model must be used for this study, due to the importance of the wave-particle resonances with all species (thermal ions, thermal electrons, and energetic ions). The multi-scale nature of the problem demands a global description, because different modes have different spatial size and localization.

In this paper, we have adopted a global gyrokinetic model and applied it to a simplified tokamak equilibrium, where numerical simulations are sufficiently fast to allow the study of the nonlinear dynamics of AMs in the presence of turbulence. 
The heat flux of AMs has been studied and compared with that of ITGs. It has been found that, in the selected regime, AMs drive an important electron heat flux, in contrast to ITGs which drive a dominant ion heat flux. This has been found to depend not only on the difference of the frequency, but also on the different spatial structure, by means of analytical estimations.

For relatively low concentration of EPs (1\%), with 10 times higher temperature than the thermal species, the radial electric field of BAEs has been observed to grow and saturate at levels one order of magnitude higher than that of ITG turbulence. As a consequence, zonal flows driven by the BAEs via forced-driven excitation have also been observed, at levels 10 times higher than those driven by turbulence alone.
The BAEs dominate the heat flux spectra at the low toroidal mode numbers and the main harmonics, injecting energy at the large spatial scales.

The global character of the problem has been studied, and we have found that the location of the BAE influences also the dynamics of turbulence in the part of the spectrum of high toroidal mode numbers. In particular, at the radial location of the BAEs, the electron heat flux becomes dominant in the presence of EPs, confirming the importance of the Alfv\'enic activity, whereas at radial locations far from the BAEs, the ITGs is found to dominate the heat flux even in the presence of EPs, and the ion heat flux remain the dominant one.

In summary, we have shown that an AM like a BAE can drive a heat flux of the thermal species as efficiently and even more efficiently than ITG turbulence, especially for the electrons. This modifies the temperature profiles of the thermal species, flattening them at the location of the AMs. The spectra of the heat fluxes are modified even in the part of high toroidal mode numbers, dominated by the ITGs, with the electron heat flux increasing in amplitude.

As next steps, we will first investigate how changing the location of the AMs influences the turbulence dynamics, in monotonic safety factor profiles and reversed-shear safety-factor profiles. In particular, one effort will be to isolate the effect of the zonal flows. These are  known to suppress the turbulence by enhancing the cascade of energy from large spatial scale to small spatial scales, and it will be important to assess how this can enter the multi-scale interaction of AMs and turbulence.
Having demonstrated the feasibility of global self-consistent gyrokinetic simulations in simplified equilibria, we will then relax some of the limitations, to approach cases which are closer to experimentally relevant scenarios.
As an ultimate goal, comprehensive theoretical studies of burning plasmas can be envisioned,  to develop a predictive capability for ITER and future fusion power plants.

\section*{Acknowledgments}
Interesting discussions with F. Zonca, Z. Qiu, E. Poli, T. G\"orler, S. Brunner, B. McMillan, E. Lanti, N. Ohana are gratefully acknowledged. This work has been carried out within the framework of the EUROfusion Consortium and has received funding from the Euratom research and training program 2014-2018 and 2019-2020 under grant agreement No 633053 within the framework of the {\emph{Multiscale Energetic particle Transport}} (MET) European Eurofusion Project. The views and opinions expressed herein do not necessarily reflect those of the European Commission. Simulations were performed on the HPC-Marconi supercomputer within the framework of the ORBFAST and OrbZONE projects. Part of this work has been done within the LABEX Plas@par project, and received financial state aid managed by the Agence Nationale de la Recherche, as part of the programme ``Investissements d'avenir'' under the reference
ANR-11-IDEX-0004-02.

\appendix

\section{Electron integrals\label{sec:Electron-integrals}}

We present the details of the evaluation of Eqs. \eqref{eq:chiN}-\eqref{eq:chiT}-\eqref{eq:chiC}.

\begin{equation}
\begin{split} & \chi_{e,T}=\pi k_{\alpha}^{2}T_{e}\int_{0}^{\infty}d\hat{v}2\pi\hat{v}^{4}\left(\hat{v}^{2}-\frac{3}{2}\right)\frac{e^{-\hat{v}^{2}}}{\sqrt{\pi}}\int_{tr}d\lambda\tau_{E}(\lambda)\left|\overline{\phi}(\lambda)\right|^{2}\delta\left(\omega_{r}-\bar{\omega}_{de}(\lambda)\right)\\
 & =\pi k_{\alpha}^{2}T_{e}\frac{\sqrt{2\epsilon}}{8}\int_{0}^{\infty}d\hat{v}2\pi\hat{v}^{2}\left(\hat{v}^{2}-\frac{3}{2}\right)\frac{e^{-\hat{v}^{2}}}{\sqrt{\pi}}\int_{0}^{1}d\kappa^{2}K\left(\kappa\right)\left|\overline{\phi}(\kappa)\right|^{2}\sum_{i}\frac{\delta(\kappa^{2}-\kappa_{i}^{2}(\hat{v}^{2}))}{-\omega_{de}^{(0)}F^{\prime}\left(\kappa_{i}^{2}(\hat{v}^{2})\right)},
\end{split}
\label{eq:elint}
\end{equation}
where $\kappa_{i}(\hat{v}^{2})$ are s.t. $\hat{v}^{2}F(\kappa_{i})=\omega_{r}/\omega_{de}^{(0)},$
with $\omega_{de}^{(0)}=m_{e}qv_{the}^{2}/(eBrR),$ and $F$ was calculated
in Ref.~\cite{Kadomtsev66}
\begin{equation}
F(\kappa)=\frac{E(\kappa)}{K(\kappa)}-\frac{1}{2}+2\hat{s}\left[\frac{E(\kappa)}{K(\kappa)}+\kappa-1\right].
\end{equation}
We will consider $\hat{s}=0,$ for simplicity. We are also neglecting
finite $\beta^{\prime}$ effects calculated in Ref.~\cite{Connor83}. Similarly, 
\begin{equation}
\begin{split} & \chi_{e,C}=\pi ek_{\alpha}\int_{0}^{\infty}d\hat{v}2\pi\hat{v}^{4}\frac{e^{-\hat{v}^{2}}}{\sqrt{\pi}}\int_{tr}d\lambda\tau_{E}\left(\lambda\right)\left|\overline{\phi}(\lambda)\right|^{2}\bar{\omega}_{de}(\lambda)\delta\left(\omega_{r}-\bar{\omega}_{de}(\lambda)\right)\\
 & =\pi ek_{\alpha}\frac{\sqrt{2\epsilon}}{8}\int_{0}^{\infty}d\hat{v}2\pi\hat{v}^{4}\frac{e^{-\hat{v}^{2}}}{\sqrt{\pi}}\int_{0}^{1}d\kappa^{2}K(\kappa)\left|\overline{\phi}(\kappa)\right|^{2}F(\kappa)\sum_{i}\frac{\delta(\kappa^{2}-\kappa_{i}^{2})}{-F^{\prime}\left(\kappa_{i}^{2}\right)}.
\end{split}
\end{equation}
Here, 

\begin{equation}
\begin{split} & \tau_{E}(\lambda)=\int_{Tr}\frac{d\theta}{\sqrt{1-\lambda B}}=\int_{Tr}\frac{d\theta}{\sqrt{1-\lambda B_{0}+\epsilon\lambda B_{0}(1-2\sin^{2}\theta/2)}}\\
 & =\int_{Tr}\frac{d\theta}{\sqrt{2\epsilon\lambda B_{0}}\sqrt{\underset{\kappa^{2}}{\underbrace{\frac{1-\lambda B_{0}}{2\epsilon\lambda B_{0}}+\frac{1}{2}}}-\sin^{2}\theta/2}}\\
 & =4\int_{0}^{2\arcsin\kappa}\frac{d\theta}{\sqrt{2\epsilon\lambda B_{0}}\sqrt{\kappa^{2}-\sin^{2}\theta/2}}\\
 & \approx\frac{4}{\sqrt{2\epsilon}}\int_{0}^{2\arcsin\kappa}\frac{d\theta}{\sqrt{\kappa^{2}-\sin^{2}\theta/2}}\\
 & =\frac{4}{\sqrt{2\epsilon}}\int_{0}^{\pi/2}\frac{2d\phi}{\sqrt{1-\kappa^{2}\sin^{2}\phi}}\\
 & \equiv\frac{8}{\sqrt{2\epsilon}}K(\kappa^{2}),
\end{split}
\end{equation}
Where $K$ is a complete elliptic integral. The bounce-angle is then
defined by $1+\cos\theta_{b}=2\kappa^{2}.$ 

Notice that for $\omega_{r}\gg\omega_{de}^{(0)},$ only the energetic
trapped electrons, $\hat{v}^{2}\gg1,$ satisfy the resonant condition
$\hat{v}^{2}F(\kappa_{i})=\omega_{r}/\omega_{de}^{(0)}.$ For a diamagnetic
mode $\omega_{r}\sim\omega_{de}^{(0)},$ and the resonance can occur
for $\hat{v}^{2}\sim1.$ 

\begin{figure}
\includegraphics[scale=0.7]{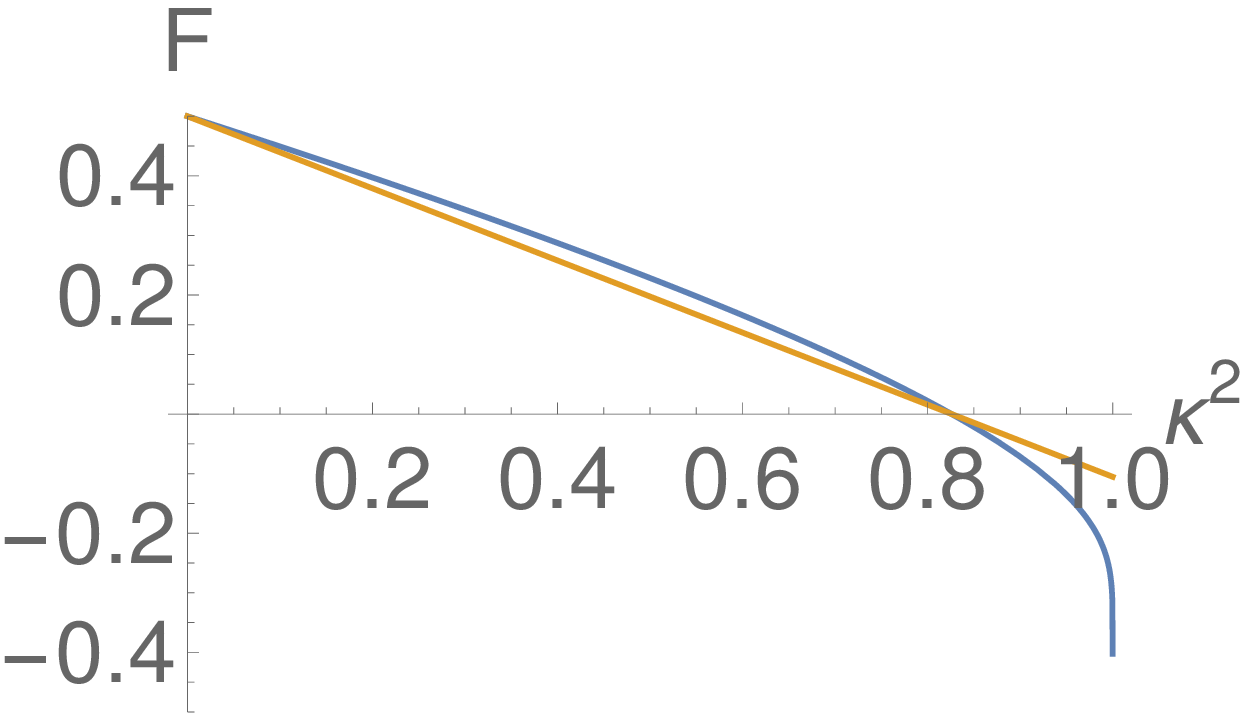}

\caption{$F$ for $\hat{s}=0.$ The line is the approximation $F=-1/2(\kappa/\kappa_{0}-1),$
with $F(\kappa_{0})=0.$}
\label{fig:fig1}
\end{figure}

For a mode rotating in the electron direction $\omega_{r}\omega_{de}^{(0)}>0,$
the condition 
\begin{equation}
\hat{v}^{2}F=\frac{\omega_{r}}{\omega_{de}^{(0)}}
\end{equation}
can be satisfied for $F$ positive, that is $0<F<1/2.$ Then, for
all values of energy below $\hat{v}_{min}^{2}=2\omega_{r}/\omega_{de}^{(0)},$
the result of the integration is zero. When $0<F<1/2,$ a very good
model is
\begin{equation}
F=-\frac{1}{2}(\kappa/\kappa_{0}-1),\,\,F(\kappa_{0})=0,
\end{equation}
see Fig. \eqref{fig:fig1} 
. Then
\begin{equation}
\kappa_{i}(\hat{v}^{2})=\kappa_{0}\left(1-2\frac{\omega_{r}}{\omega_{de}^{(0)}\hat{v}^{2}}\right),
\end{equation}
with $\kappa_{0}=0.826225,$ and $0\leq\kappa_{i}(\hat{v}^{2})\leq\kappa_{0},$
for any $\hat{v}_{min}\leq\hat{v}<\infty.$ Thus, 
\begin{equation}
-F^{\prime}\left(\kappa_{i}(\hat{v}^{2})\right)=\frac{1}{2\kappa_{0}},\,\mbox{for}\,\hat{v}_{min}\leq\hat{v}<\infty.
\end{equation}

For a mode rotating in the ion direction, $\omega_{r}/\omega_{de}^{(0)}<0,$
we must consider negative values of $F,$ that is $\kappa_{0}\leq\kappa<1.$

\begin{figure}
\includegraphics[scale=0.7]{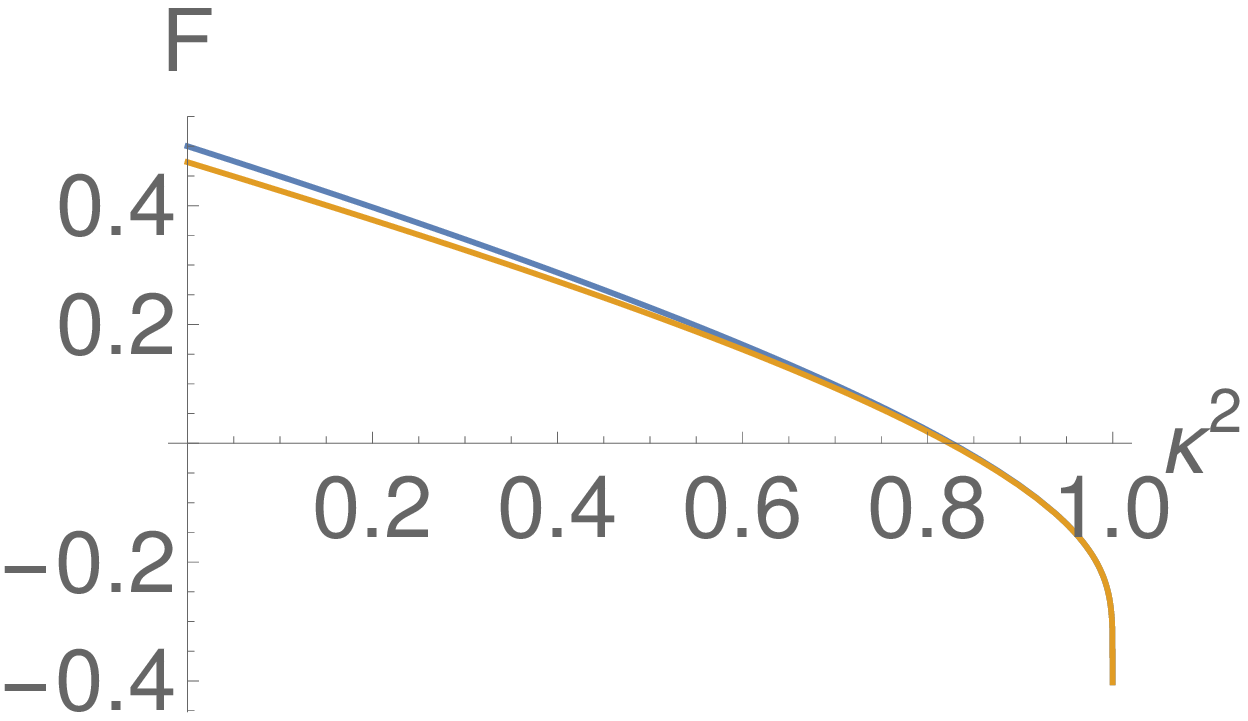}

\caption{$F,$ full and approximated according to Eq. \eqref{eq:apprF}}
\label{fig:fig2}
\end{figure}

A very good approximation that we found (by simply Taylor expanding
around $\kappa=1^{-}$) is the following one {[}see Fig \eqref{fig:fig2}{]}

\begin{figure}
\includegraphics[scale=0.7]{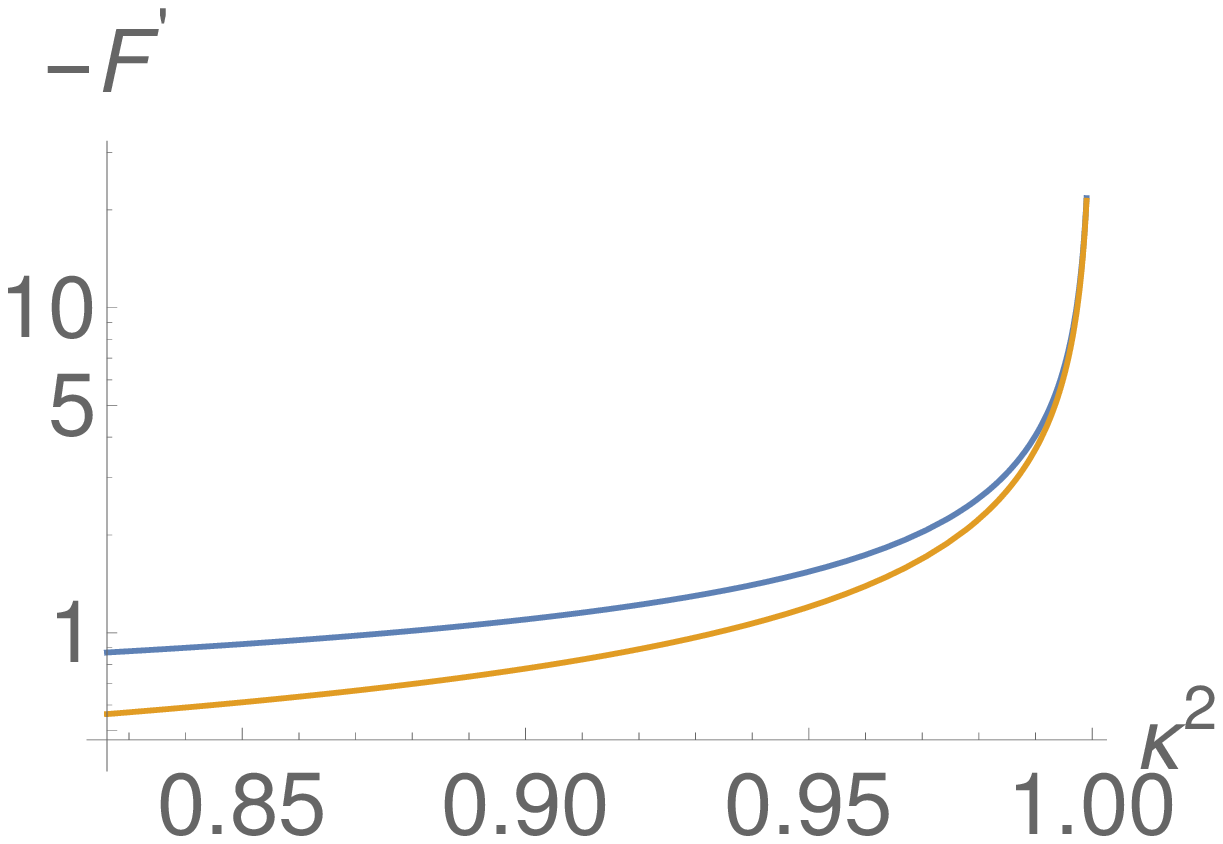}

\caption{$F^{\prime},$ full and approximated according to Eq. \eqref{eq:FprimA}}
\label{fig:fig3}
\end{figure}

\begin{equation}
\begin{split} & F\approx\frac{2\left(7-2\log2-\kappa(6\log2-1)\right)+\left(1+3\kappa\right)\log\left(1\text{-\ensuremath{\kappa}}\right)}{-4(1-10\log2)-4\kappa\left(-1+2\log2\right)-2\left(5-\kappa\right)\log\left(1-\kappa\right)}\\
 & \equiv\frac{a+b\kappa+\left(1+3\kappa\right)\log\left(1\text{-\ensuremath{\kappa}}\right)}{c+d\kappa-2\left(5-\kappa\right)\log\left(1\text{-\ensuremath{\kappa}}\right)},
\end{split}
\label{eq:apprF}
\end{equation}
definitely valid for $-1/2\leq F\leq0,$ and beyond!
For $\kappa_{0}<\kappa \lesssim 1,$
we then find
\begin{equation}
\kappa_{i}\left(\hat{v}^{2}\right)\approx1-e^{-\frac{1}{4}\frac{\left(a+b\right)\hat{v}^{2}-\left(c+d\right)\frac{\omega_{r}}{\omega_{de}^{(0)}}}{\hat{v}^{2}+2\frac{\omega_{r}}{\omega_{de}^{(0)}}}},
\end{equation}
where, again $\hat{v}_{min}\leq\hat{v}<\infty,$ with $\hat{v}_{min}=\sqrt{2\left|\omega_{r}/\omega_{de}^{(0)}\right|}.$
Therefore
\begin{equation}
F^{\prime}\left(\kappa_{i}(\hat{v}^{2})\right)\approx4\frac{2(a+b)+c+d}{\left[1-\kappa_{i}(\hat{v}^{2})\right]\left\{ c+d-8\log\left(1-\kappa_{i}(\hat{v}^{2})\right)\right\} ^{2}}.\label{eq:FprimA}
\end{equation}
This approximation is also quite good {[}see Fig \ref{fig:fig3}{]}


\begin{thebibliography}{99}

\bibitem{Rudakov65} L. I. Rudakov, and R. Z. Sagdeev, {\it Sov. Phys. Dokl.} {\bf 6} 498 (1965)

\bibitem{Hasegawa79} A. Hasegawa, C.G. Maclennan and Y. Kodama, {\it Phys. Fluids} {\bf 22}, 2122 (1979)

\bibitem{Appert82} K. Appert, R. Gruber, F. Troyon, J. Vaclavik, {\it Plasma Phys. Nucl. Fusion} {\bf 24}, 1147 (1982)

\bibitem{Cheng85} C. Z. Cheng and M. S. Chance, {\it Ann. Phys.} {\bf 161}, 21  (1985)

\bibitem{Chu92} M. S. Chu, J. M. Greene, L. L. Lao, A. D. Turnbull, M. S. Chance, {\it Phys. Fluids} B {\bf 4}, 3713  (1992)

\bibitem{Heibrink93} W. W. Heidbrink, E. J. Strait, M. S. Chu, and A. D. Turnbull, {\it Phys. Rev. Lett.} {\bf 71}, 855 (1993)

\bibitem{Zonca96} F. Zonca, L. Chen and R.A. Santoro {\it Plasma Ph. Control. Fusion} {\bf 38}, 2011-2028 (1996) 

\bibitem{Chen16} L. Chen and F. Zonca, {\it Rev. Mod. Phys.} {\bf 88}, 015008 (2016)

\bibitem{Maraschek97} M. Maraschek, S. G\"unter, T. Kass, B. Scott, H. Zohm, and ASDEX Upgrade Team, {\it Phys. Rev. Lett.} {\bf 79}, 4186 (1997)

\bibitem{Chen98} L. Chen, F. Zonca, R. A. Santoro, and G. Hu {\it Plasma Phys. Control. Fusion} {\bf 40}, 1823 (1998)




\bibitem{Ida20} K. Ida, {\it Plasma Phys. Control. Fusion} {\bf 62}, 014008 (2020)

\bibitem{Chen00} L. Chen, Z. Lin, and R. White,  {\it Phys. Plasmas} {\bf 7}, 3129 (2000)

\bibitem{Chen12} L. Chen, and F. Zonca, {\it Phys. Rev. Lett.} {\bf 109}, 145002 (2012)

\bibitem{Qiu16PoP} Z. Qiu, L. Chen, and F. Zonca, {\it Phys. Plasmas} {\bf 23}, 090702 (2016)

\bibitem{Qiu16NuFu} Z. Qiu, L. Chen, and F. Zonca, {\it Nucl. Fusion} {\bf 56}, 106013 (2016)




\bibitem{Zonca15} F. Zonca, L. Chen, S. Briguglio, G. Fogaccia, A. V. Milovanov, Z. Qiu, G. Vlad, and X. Wang, {\it Plasma Phys. Control. Fusion} {\bf 57} 014024 (2015)

\bibitem{Diamond05} P. H. Diamond, S. I. Itoh, K. Itoh, and T. S. Hahm, {\it Plasma Phys. Control. Fusion} {\bf 47}, R35 (2005)

\bibitem{Chen07} L. Chen and F. Zonca, {\it Nucl. Fusion} {\bf 47}, 886 (2007)

\bibitem{Falessi19} M. V. Falessi and F. Zonca, {\it Phys. Plasmas} {\bf 26} 022305 (2019)


\bibitem{Tardini07} G. Tardini, et al., {\it Nucl. Fusion} {\bf 47}, 280 (2007)




\bibitem{Heidbrink09} W. W. Heidbrink, J. M. Park, M. Murakami, C. C. Petty, C. Holcomb, and M. A. Van Zeeland, {\it Phys. Rev. Lett.} {\bf 103}, 175001 (2009)

\bibitem{Romanelli10} M. Romanelli, A. Zocco, F. Crisanti, and JET Contributors, {\it Plasma Phys. Control. Fusion} {\bf 52}, 045007 (2010

\bibitem{Bock17} A. Bock, et al., {\it Nucl. Fusion} {\bf 57} 126041 (2017)

\bibitem{White89} R. White, and H. Mynick, {\it Phys. Fluids B} {\bf 1}, 980 (1989)

\bibitem{Angioni09} C. Angioni, A. G. Peeters, G. V. Pereverzev, A. Bottino, J. Candy, R. Dux, E. Fable, T. Hein, and R. E. Waltz, {\it Nucl. Fusion} {\bf 49}, 055013 (2009)

\bibitem{Zhang10} W. Zhang, V. Decyk, I. Holod, Y. Xiao, Z. Lin, and L. Chen {\it Phys. Plasmas} {\bf 17}, 055902 (2010)

\bibitem{Holland12} C. Holland, C. C. Petty, L. Schmitz, K. H. Burrell, G. R. McKee, T. L. Rhodes, and J. Candy, {\it Nucl. Fusion} {\bf 52}, 114007 (2012)

\bibitem{Citrin13}  J. Citrin, et al, {\it Phys. Rev. Lett.} {\bf 111}, 155001 (2013)

\bibitem{Garcia15} J. Garcia, C. Challis, J. Citrin, H. Doerk, G. Giruzzi, T. G\"orler,
F. Jenko, P. Maget, and JET Contributors, {\it Nucl. Fusion} {\bf 55} 053007 (2015)

\bibitem{DiSiena18} A. Di Siena, T. G\"orler, H. Doerk, E. Poli, and R. Bilato, {\it Nucl. Fusion} {\bf 58} 054002 (2018)





\bibitem{DiSiena19} A. Di Siena, T. G\"orler, E. Poli, A. Ba\~non Navarro, A. Biancalani, and F. Jenko, {\it Nucl. Fusion} {\bf 59}, 124001 (2019)

\bibitem{Bass10} E. M. Bass and R. E. Waltz, {\it Phys. Plasmas} {\bf 17}, 112319 (2010)

\bibitem{Cole17} M. D. J. Cole, A. Biancalani, A. Bottino, R. Kleiber, A. K\"onies, A. Mishchenko, {\it Phys. Plasmas} {\bf 24}, 022508 (2017)

\bibitem{Biancalani20JPP} A. Biancalani, A. Bottino, P. Lauber, A. Mishchenko, and F. Vannini, {\it J. Plasma Phys.} {\bf 86}, 825860301  (2020)

\bibitem{Jolliet07} S. Jolliet, A. Bottino, P. Angelino, R. Hatzky, T. M. Tran, B. F. Mcmillan, O. Sauter, K. Appert, Y. Idomura, and L. Villard, {\it Comput. Phys. Commun.} {\bf 177}, 409 (2007)

\bibitem{Bottino11} A. Bottino, T. Vernay, B. Scott, S. Brunner, R. Hatzky, S. Jolliet, B. F. McMillan, T. M. Tran, and L. Villard, {\it Plasma Phys. Control. Fusion} {\bf 53}, 124027 (2011)

\bibitem{Lanti20} E. Lanti, et al, {\it Comp. Phys. Commun.} {\bf 251}, 107072 (2020)

\bibitem{Tronko19} N. Tronko, A. Bottino, C. Chandre, E. Sonnendr\"ucker, S. Brunner, E. Lanti, N. Ohana, and L. Villard, {\it Plasma Phys. Control. Fusion} {\bf 61}, 114002 (2019)

\bibitem{Koenies18} A. K\"onies, et al, {\it Nucl. Fusion} {\bf 58} 12, 126027 (2018)

\bibitem{Taimourzadeh19} S. Taimourzadeh, et al, {\it Nucl. Fusion} {\bf 59}, 066006 (2019)




\bibitem{Biancalani14} A. Biancalani, A. Bottino, Ph. Lauber, D. Zarzoso, {\it Nucl. Fusion} {\bf 54}, 104004 (2014)

\bibitem{Biancalani17pop} A. Biancalani, et al, {\it Phys. Plasmas} {\bf 24}, 062512 (2017)

\bibitem{Goerler16} T. G\"orler, N. Tronko, W. A. Hornsby, A. Bottino, R. Kleiber, C. Norscini, V. Grandgirard, F. Jenko, and E. Sonnendr\"ucker, {\it  Phys. Plasmas} {\bf 23}, 072503 (2016)

\bibitem{Tronko17} N. Tronko,  A. Bottino, T. G\"orler, E. Sonnendr\"ucker, D. Told, and L. Villard, {\it Phys. Plasmas} {\bf 24}, 056115 (2017)

\bibitem{Biancalani19EPS} A. Biancalani, et al., {\it EPS Conference, Division of Plasma Physics}, July 2019, Milan, Italy, I5.J602, \url{http://ocs.ciemat.es/EPS2019ABS/pdf/I5.J602.pdf}

\bibitem{Mishchenko17} A. Mishchenko, A. Bottino, R. Hatzky, E. Sonnendr\"ucker, R. Kleiber, and A. K\"onies, {\it Phys. Plasmas} {\bf 24}, 081206 (2017)

\bibitem{Mishchenko18} A. Mishchenko, A. Bottino, A. Biancalani, R. Hatzky, T. Hayward-Schneider, N. Ohana, E. Lanti, S. Brunner, L. Villard, M. Borchardt, R. Kleiber, A. K\"onies, {\it Comp. Phys. Comm.} {\bf 238}, 194 (2018)

\bibitem{Novikau20} I. Novikau, et al, {\it Phys. Plasmas} {\bf 27}, 042512 (2020)

\bibitem{Vannini20} F. Vannini, A. Biancalani, A. Bottino, T. Hayward-Schneider, Ph. Lauber, A. Mishchenko, I. Novikau, E. Poli and the ASDEX Upgrade team, {\it Phys. Plasmas} {\bf 27}, 042501 (2020)

\bibitem{Novikau19} I. Novikau, A. Biancalani, A. Bottino, A. Di Siena, P. Lauber, E. Poli, E. Lanti, L. Villard, N. Ohana, and S. Briguglio, {\it accepted for publication in Comput. Phys. Commun.} (2019)  \url{https://www.sciencedirect.com/science/article/pii/S0010465519303704}





\bibitem{Briguglio00} S. Briguglio, L. Chen, Jiaqi Dong, G. Fogaccia, R.A. Santoro, G. Vlad, and F. Zonca, {\it Nucl. Fusion} {\bf 40} 701 (2000) 

\bibitem{Chen99} L. Chen, {\it J. Geophys. Res.} {\bf 104}, 2421 (1999)

\bibitem{Stutman09} D. Stutman, L. Delgado-Aparicio, N. Gorelenkov, M. Finkenthal, E. Fredrickson, S. Kaye, E. Mazzucato, and K. Tritz, {\it Phys. Rev. Lett.} {\bf 102}, 115002 (2009)

\bibitem{Hayward19} T. Hayward-Schneider, PhD thesis (2020)

\bibitem{Zocco15} A. Zocco, P. Helander, and J. W. Connor {\it Plasma Ph. Control. Fusion} {\bf 57}, 085003 (2015)

\bibitem{Kadomtsev66} B. B. Kadomtsev, and O. P. Pogutse, {\it Soviet Phys. JETP} {\bf 24}, 1172 (1966)

\bibitem{Connor83} J.W. Connor, R.J. Hastie, T.J. Martin {\it Nucl. Fusion} {\bf 23}, 1702 (1983)








\end{thebibliography}
\end{document}